\documentclass[12pt]{article}
\usepackage{graphicx}
\usepackage{amssymb,amsmath,amsfonts}

\def\Re{{\cal R \mskip-4mu \lower.1ex \hbox{\it e}\,}}
\def\Im{{\cal I \mskip-5mu \lower.1ex \hbox{\it m}\,}}
\def\ie{{\it i.e.}}
\def\eg{{\it e.g.}}

\def\etal{{\it et al.}}

\def\sub#1{_{\lower.25ex\hbox{$\scriptstyle#1$}}}
\def\tev{\,{\ifmmode\mathrm {TeV}\else TeV\fi}}
\def\gev{\,{\ifmmode\mathrm {GeV}\else GeV\fi}}
\def\mev{\,{\ifmmode\mathrm {MeV}\else MeV\fi}}
\def\mpl{\ifmmode M_{pl}\else $M_{pl}$\fi}
\def\mpl{\ifmmode \overline M_{Pl}\else $\bar M_{Pl}$\fi}
\def\to{\rightarrow}

\def\subw{_{\rm w}}
\def\mh{\ifmmode m\sbl H \else $m\sbl H$\fi}
\def\mch{\ifmmode m_{H^\pm} \else $m_{H^\pm}$\fi}
\def\mt{\ifmmode m_t\else $m_t$\fi}
\def\mc{\ifmmode m_c\else $m_c$\fi}
\def\mz{\ifmmode M_Z\else $M_Z$\fi}
\def\mw{\ifmmode M_W\else $M_W$\fi}
\def\mws{\ifmmode M_W^2 \else $M_W^2$\fi}
\def\mhs{\ifmmode m_H^2 \else $m_H^2$\fi}   
\def\mzs{\ifmmode M_Z^2 \else $M_Z^2$\fi}
\def\mts{\ifmmode m_t^2 \else $m_t^2$\fi}
\def\mcs{\ifmmode m_c^2 \else $m_c^2$\fi}
\def\mchs{\ifmmode m_{H^\pm}^2 \else $m_{H^\pm}^2$\fi}
\def\ztwo{\ifmmode Z_2\else $Z_2$\fi}
\def\zone{\ifmmode Z_1\else $Z_1$\fi}
\def\mtwo{\ifmmode M_2\else $M_2$\fi}
\def\mone{\ifmmode M_1\else $M_1$\fi}
\def\tb{\ifmmode \tan\beta \else $\tan\beta$\fi}
\def\xw{\ifmmode x\subw\else $x\subw$\fi}
\def\ch{\ifmmode H^\pm \else $H^\pm$\fi}
\def\lum{\ifmmode {\cal L}\else ${\cal L}$\fi}
\def\inpb{\,{\ifmmode {\mathrm {pb}}^{-1}\else ${\mathrm {pb}}^{-1}$\fi}}
\def\infb{\,{\ifmmode {\mathrm {fb}}^{-1}\else ${\mathrm {fb}}^{-1}$\fi}}
\def\epem{\ifmmode e^+e^-\else $e^+e^-$\fi}
\def\ppb{\ifmmode \bar pp\else $\bar pp$\fi}
\def\bsg{\ifmmode B\to X_s\gamma\else $B\to X_s\gamma$\fi}
\def\bsll{\ifmmode B\to X_s\ell^+\ell^-\else $B\to X_s\ell^+\ell^-$\fi}
\def\bstt{\ifmmode B\to X_s\tau^+\tau^-\else $B\to X_s\tau^+\tau^-$\fi}
\def\lamt{\ifmmode \tilde\lambda\else $\tilde\lambda$\fi}
\def\shat{\ifmmode \hat s\else $\hat s$\fi}
\def\that{\ifmmode \hat t\else $\hat t$\fi}
\def\uhat{\ifmmode \hat u\else $\hat u$\fi}

\newskip\zatskip \zatskip=0pt plus0pt minus0pt
\def\matth{\mathsurround=0pt}
\def\lsim{\mathrel{\mathpalette\atversim<}}
\def\gsim{\mathrel{\mathpalette\atversim>}}
\def\atversim#1#2{\lower0.7ex\vbox{\baselineskip\zatskip\lineskip\zatskip
  \lineskiplimit 0pt\ialign{$\matth#1\hfil##\hfil$\crcr#2\crcr\sim\crcr}}}

\def\grtsim{\,\,\rlap{\raise 3pt\hbox{$>$}}{\lower 3pt\hbox{$\sim$}}\,\,}
\def\lsim{\,\,\rlap{\raise 3pt\hbox{$<$}}{\lower 3pt\hbox{$\sim$}}\,\,}

\renewcommand{\thefootnote}{\fnsymbol{footnote}}

\begin{document} \begin{titlepage}
\rightline{\vbox{\halign{&#\hfil\cr
&SLAC-PUB-12689\cr
}}}
\begin{center}
\thispagestyle{empty} \flushbottom { {
\Large\bf Collider Signals of Gravitational Fixed Points
\footnote{Work supported in part
by the Department of Energy, Contract DE-AC02-76SF00515}
\footnote{e-mail:
$^a$hewett@slac.stanford.edu,$^b$rizzo@slac.stanford.edu}}}
\medskip
\end{center}

\centerline{JoAnne L. Hewett$^{a}$ and Thomas G. Rizzo$^{b}$}
\vspace{8pt} 
\centerline{\it Stanford Linear
Accelerator Center, 2575 Sand Hill Rd., Menlo Park, CA, 94025}

\vspace*{0.3cm}

\begin{abstract}

Recent studies have shown that the poor perturbative behavior of
General Relativity in the ultraviolet regime may be ameliorated by the
existence of a non-Gaussian fixed point which renders the theory
asymptotically safe and possibly non-perturbatively renormalizable.  This
results in a running of the (effective) gravitational coupling such 
that gravity becomes weaker at high energies.  We parameterize this 
effective coupling with a form factor and study its consequences at the 
LHC and ILC in models with large extra dimensions or warped extra 
dimensions.  We find significant effects in the processes of Kaluza-Klein 
(KK) graviton exchange or resonant KK graviton production in both the 
Drell-Yan reaction as well as in $e^+e^-\to f\bar f$. On the otherhand, 
processes leading to KK graviton emission show qualitatively less 
sensitivity to the presence of a form factor.  In addition, we examine 
tree-level perturbative unitarity in $2\to 2$ gravity-mediated scattering 
and find that this form factor produces a far better behaved amplitude 
at large center of mass energies.

\end{abstract}



\renewcommand{\thefootnote}{\arabic{footnote}} \end{titlepage} 

%
%
%

\section{Introduction}

Over the past 20 years we have come to learn that all strong 
and electroweak phenomena below the scale of a few hundred GeV 
can be well described \cite{sm} with high precision by renormalizable 
Yang-Mills gauge theories as encoded in the Standard 
Model (SM) Lagrangian. Of course within this realm some issues 
remain to be addressed, such as the origin of electroweak 
symmetry breaking (which we hope to resolve from data provided 
by the LHC) and the origin of the fermion family structure. 
Similarly, the description of gravity via Einstein's General 
Relativity (GR), as encoded in the Einstein-Hilbert (EH) action, 
has been proven remarkably successful over a wide range of scales 
from the sub-millimeter \cite{adel}, to interplanetary \cite{solar}, and 
even cosmological distances \cite{cosmo}. Here, too, some issues remain 
to be addressed such as the nature of the (apparent) 
dark energy/cosmological constant. 
The next broad question to answer is how to unify our description 
of gravity with those of the other 
three forces, \ie, how do we construct a quantum theory of gravity. 
This problem is long-standing and has been the subject of 
much labor over the last half-century \cite{Feynman:1963ax}, and our 
perspective on possible ways to resolve this problem have 
evolved significantly over time.

The essential issue with constructing quantum gravity in the 
standard approach is that the EH action leads to a quantum field 
theory which is not perturbatively renormalizable, unlike the 
case of Yang-Mills theories. This can be  most easily seen by 
examining the interaction of gravitons with matter (or their 
self-interactions) in any fixed metric background. Distinct from the 
case of Yang-Mills theories, where the interactions of gauge fields with 
matter or each other correspond to dimension-4 operators (in 4-dimensional
spacetime), 
the gravitational interactions correspond to operators of 
dimension-5 (or higher) leading to non-renormalizability by simple 
power counting.  This implies that the theory is not well-behaved 
in the ultraviolet (UV). 
In particular, this approach implies that (4-d) gravity 
becomes strong near the reduced 
Planck scale, $\overline M_{Pl}$, so that we usually 
treat GR as an effective theory at energies far below that 
scale. As we have been recently reminded \cite{Ward:2006bc}, 
Weinberg \cite{Weinberg} long ago pointed out that there are only 
four known  
approaches to dealing with this issue, and they have not evolved 
qualitatively since that time:
\begin{itemize}

\item {The EH action in 4 dimensions is incomplete; new physics must 
be added that somehow tames the poor perturbative UV behavior of 
GR. This is the path followed by String Theory (where extra dimensions, 
supersymmetry, new fields and a new scale enter) \cite{strings}  
as well as in Loop Quantum Gravity (which also introduces a new scale)
\cite{Smolin:2004sx}. This approach has received the most attention 
in recent years and has met with a number of successes.}

\item {Gravitons are not fundamental objects but are composite; 
this possibility has been addressed by a number of 
authors \cite{compgrav}.}

\item {The poor UV behavior of GR in 4-d can be controlled by a 
re-ordering of the conventional perturbation expansion
employing a modified version of the Yennie, Frautschi and 
Suura \cite{Yennie:1961ad} resummation techniques. This approach has 
recently been advocated by Ward \cite{Ward:2006bc}}.

\item {The poor UV behavior of GR is an artifact of perturbation 
theory. General Relativity is non-perturbatively renormalizable having the 
property of being asymptotically safe due to 
the existence of a non-Gaussian fixed point; such an approach 
has also met with a number of recent 
successes \cite{bigref}. The existence of such a fixed point
has been demonstrated in both 
field theoretical and in lattice studies \cite{Hamber:2006sv} in 4-d as 
well as in higher dimensions \cite{Fischer:2006at}. 
In such an approach, the running gravitational coupling becomes 
weaker as the fundamental gravity scale is reached.}

\end{itemize}

A common feature of the last two approaches is that the strength 
of the gravitational interaction, usually expressed through 
Newton's constant, $G_N$, runs in such a way that the effective 
coupling actually becomes {\it weaker} in the energy 
regime near $\overline M_{Pl}$. 
This would imply good high energy behavior and, perhaps, 
the restoration of unitarity in graviton scattering amplitudes. 
Could such ideas be tested in, \eg, collider experiments? Clearly, 
to do so collision energies must 
approach the fundamental scale of gravity, which is unattainable 
4-dimensional gravity. However, the property of asymptotic safety has been 
demonstrated to persist in higher dimensions \cite{Fischer:2006at} where 
we know that we can construct scenarios where the (true) fundamental 
scale of gravity is of order $\sim$ TeV, such as in the models of 
Arkani-Hamed, Dimopoulos and Dvali(ADD) {\cite {ADD}} with large extra
dimensions and of Randall and 
Sundrum(RS) {\cite {RS}} with warped geometries. 
In this paper, we will demonstrate 
that if either the ADD or RS models are realized they, will provide a  
framework for testing the hypothesis of asymptotic safety at 
the LHC and ILC through the appearance of gravitational  
form-factors which will damp the strength of gravity at high 
energies. As we will see below, such form-factors will lead to 
significant modifications in the traditional predictions for both 
of these scenarios.  These deviations from the standard predictions
can then be used to test the nature of the form factor and to determine
if a consistent theory along such lines can be successfully 
constructed.

The outline of this paper is as follows: In Section II, we will 
provide the essential background formalism for applying the 
approach of asymptotic safety to both the ADD and RS extra-dimensional 
models. In Sections III and IV we will examine how the 
traditional signals of the ADD and RS models, respectively, at 
future colliders are modified and the parameter space 
range over which they may be observed. A discussion and a summary 
of our results can be found in Section V.

\section{Background Formalism}

Here we present the formalism that is relevant to our
analysis.  A general review of non-perturbative renormalizability 
and asymptotic safety is clearly beyond the scope of the present 
work, but can be found in the introductory survey by Niedermaier 
\cite{Niedermaier:2006ns}.
We work in Euclidean space within the context of an effective 
average action that consists of a truncated list of operators 
which are part of a more generalized gravitational action. The simplest
choice for this fixed set of operators is the 
Einstein-Hilbert truncation.  This action is simply given by the 
familiar expression in $D$-dimensions,
\begin{equation}
S_{EH}=\int d^Dx ~\sqrt {-g} \big[{M^{D-2}\over {2}}~R-\Lambda\big]\,,
\end{equation}
where $M$ is the $D$-dimensional Planck scale, $R$ is the Ricci scalar,
$\Lambda$ corresponds to the cosmological constant, and $D=\delta+3+1$
where $\delta$ is the number of additional spatial dimensions.{\footnote 
{For consistency with the structure of the ADD model, we drop the 
cosmological constant term from the above
action in our ensuing discussion.}} 
Newton's constant in $D$-dimensions can then be defined in analogy 
with 4-dimensional expression as $G_D=1/(8\pi M^{D-2})$.  For convenience,
we define a corresponding dimensionless quantity, 
\begin{equation}
g(\mu)=\mu^{D-2}G_D\,,
\end{equation}
where $\mu$ is an arbitrary mass scale. Our goal is to obtain the
renormalization group equations (RGEs) for this dimensionless
gravitational coupling and we are particularly 
interested in the RGE behavior of $g$ in the ultraviolet (UV).
It has been found that 
the qualitative nature of this behavior is not very sensitive 
to the truncation in the gravitational theory employed above where 
only the EH term appears in the action \cite{bigref}.  

Following Bonanno and Reuter \cite{Bonanno:2006eu}, the relevant  
RGE corresponding to the action above is found to be  
\begin{equation}
{{dg}\over {dt}}=[D-2+\eta]g\,,
\end{equation}
where $t=\log(\mu)$ and $\eta$ is the non-perturbative 
anomalous dimension of the EH operator given by 
\begin{equation}
\eta={{gB_1}\over {1-gB_2}}\,.
\end{equation}
Here, the constants $B_{1,2}$ can be expressed in terms of a 
set of integrals which effectively result from loop summations: 
\begin{eqnarray}
B_1={1\over {3}}~(4\pi)^{1-D/2} \Big[D(D-3)\phi_1-[6D(D-1)+24]
\phi_2 \Big]\nonumber \\
B_2=-{1\over {6}}~(4\pi)^{1-D/2} \Big[D(D+1)\tilde \phi_1-6D(D-1)
\tilde \phi_2 \Big]\,,
\end{eqnarray}
where the integrals are explicitly given by 
\begin{eqnarray}
\phi_1={1\over {\Gamma(D/2-1)}} \int_0^\infty ~dz~z^{D/2-2} ~{{Q-zQ'}
\over {z+Q}}\nonumber \\
\phi_2={1\over {\Gamma(D/2)}} \int_0^\infty ~dz~z^{D/2-1} ~{{Q-zQ'}
\over {(z+Q)^2}}\nonumber \\
\tilde \phi_1={1\over {\Gamma(D/2-1)}} \int_0^\infty ~dz~z^{D/2-2} ~{{Q}
\over {z+Q}}\nonumber \\
\phi_1={1\over {\Gamma(D/2)}} \int_0^\infty ~dz~z^{D/2-1} ~{{Q}
\over {(z+Q)^2}}\,.
\end{eqnarray}
Here, $Q$ is an essentially arbitrary smooth cutoff function 
with the properties $Q(0)=1$ and $Q(z)\to 0$ as $z\to \infty$, and the
derivative is defined by  
$Q'=dQ/dz$. We take $Q=z/(e^z-1)$ in explicit 
computations and define the  parameters 
$\omega=-B_1/2$ and $\omega'=\omega+B_2$ as suggested in 
\cite{Bonanno:2006eu}. 
The RGE is now seen to exhibit two fixed points where $\beta=D-2+\eta=0$: 
($i$) an attractive infrared (IR) Gaussian (or perturbative) fixed point 
at $g^{IR}=0$ and ($ii$) an UV 
attractive non-Gaussian fixed point where $g^{UV}=1/\omega'$. 

The RGE differential equation above can be solved analytically. 
Taking $\mu=\mu_0$ as a boundary condition and defining 
$g_0\equiv g(\mu_0)$ we obtain 
\begin{equation}
{g\over {(1-\omega'g)^{\omega/\omega'}}}={{g_0}\over 
{[1-\omega'g_0]^{\omega/\omega'}}}\Big({\mu\over {\mu_0}}\Big)^{D-2}\,.
\end{equation}
This solution, in itself, cannot be solved analytically for
$g(\mu)$ in closed form. 
However, a numerical analysis shows that $\omega \simeq \omega'$ to
order $\sim 10\%$; this allows
for an approximate analytical solution to be obtained which is given by  
\begin{equation}
g(\mu)\simeq {{(\mu/\mu_0)^{D-2}g_0}\over 
{[1+\omega (\mu/\mu_0)^{D-2}g_0-1]}}\,.
\end{equation}

Rewriting this result in terms of the $D$-dimensional Planck scale and
taking the limit $\mu_0\to 0$ yields the effect of this RGE evolution on
the gravitational coupling in the ADD model. 
This leads to a mapping into an effective $D$-dimensional Planck scale of
\begin{equation}
{1\over {M^{D-2}}}\to {1\over {M_{eff}^{D-2}}}={1\over {M^{D-2}}}
\Big[1+{{\omega}\over {8\pi}}({{\mu^2}\over {M^2}})
^{D/2-1}\Big]^{-1}\,,
\end{equation}
which is then used in the $D$-dimensional coupling of the graviton field 
$H^{AB}$ to the matter stress-energy tensor, $T_{AB}$, \ie, 
\begin{equation}
{\cal L}=-T_{AB}H^{AB}/M_{eff}^{D-2}\,.
\end{equation}
We can instead write this rescaling relation as 
\begin{equation}
{1\over {M^{D-2}}}\to {1\over {M^{D-2}}}F(\mu^2)\,,
\end{equation}
where $F$ can be treated as a form factor now appearing in the 
effective coupling which we can rewrite as 
\begin{equation}
F=\Big[1+\Big({{\mu^2}\over {t^2M^2}}\Big)^{1+\delta/2}\Big]^{-1}\,,
\label{formf}
\end{equation}
where $\delta=D-4$ is the number of additional dimensions. 
Numerically, we find that the parameter $t$ (which is trivially related
to $\omega$) is quite close 
to unity assuming that $5\leq D \leq 11$ as is true for all 
the cases of interest 
to us here. In our analysis below we will treat $t$ as an 
${\cal O}(1)$ free parameter to allow for uncertainties in the calculation
above which arise from, \eg, the truncation of the EH action, our specific
choice for the function $Q$, and the small violation of the $\omega\simeq
\omega'$ relation.
  
Note that this form factor ensures
that the gravitational coupling is unaffected at low energies and 
retains the value of the fundamental Planck scale, but then runs 
with increasing energy.  The derivation of this form factor has not 
relied on the background metric and hence it can be equally well applied
to the case of warped geometries as well as flat dimensions, provided
that, for simplicity, the possibility of a running cosmological 
constant is neglected.

In order to quantify the effect of this form factor in the collider
signatures of either 
the ADD or RS models, we need to relate the quantity $\mu$ to
physical parameters in the production process; this 
issue is similar to that of the apparent scale ambiguity in QCD in
computations performed at finite order in perturbation theory. 
In reactions mediated by $s$-channel kinematics, which are typical of graviton 
exchange or resonant production processes in the ADD or RS models, 
it is natural to take $\mu^2=s$ so that in such cases the
form factor becomes    
\begin{equation}
F=\Big[1+\Big({{\sqrt s}\over {tM}}\Big)^{\delta+2}\Big]^{-1}\,.
\end{equation}
Thus in the graviton exchange process in the ADD model, which is  
described by a cutoff parameter $\Lambda_H$ \cite{rev,me}, 
the corresponding result 
for the cross section including the form factor can be obtained 
by making the replacement 
\begin{equation}
\Lambda_H^4 \to \Lambda_H^4\Big[1+\Big({{\sqrt s}\over 
{tM}}\Big)^{\delta+2}\Big]\,.
\end{equation}
Note that this renders the predictions for graviton exchange explicitly
dependent on $\delta$, which does not occur within this formalism 
in the traditional ADD scenario.
Since $\Lambda_H \simeq M$ in ADD, we can also make the substitution
$tM\to t'\Lambda_H$, where $t'$ is another $O(1)$ parameter.  (In our
analyses below we will not make the distinction between $t$ and $t'$.) 
Note that for fixed $\Lambda_H$, the effect of the form factor 
increases as $t$ (or $t'$) take on smaller values. 
In other processes, such as graviton emission in ADD, there is more
ambiguity in the identification of the scale $\mu$.  Here, one may
choose from several different kinematic quantities 
such as the emitted graviton's energy or $p_T$.  This ambiguity will 
affect the numerical values of the cross sections
in detail and we will thus present results 
for several different choices for this scale.  However, as we will see
below, our qualitative results are insensitive to this uncertainty.   

We now examine the effects of this form factor in the collider signatures
for the ADD and RS models and determine whether the resulting suppression
in the strength of the gravitational coupling is observable in these
scenarios.

\section{Large Extra Dimensions}

In this scenario \cite{ADD}, the Standard Model fields are confined to a
3-brane in the higher dimensional bulk and gravity alone progrates in
the additional dimensions.  The 4-dimensional Planck scale is related
to the fundamental scale $M$ via
\begin{equation}
\overline M_{Pl}^2 = V_\delta M^{2+\delta}\,,
\end{equation}
where $V_\delta$ is the volume of the extra dimensional space.  Taking
$M\sim$ TeV eliminates the hierarchy between $\overline M_{Pl}$ and the
electroweak scale.  A vast number of studies have been performed investigating
the consequences of this framework \cite{rev}.  Here, we explore the
modifications in the high energy collider signals that are introduced 
by the presence of a running gravitational coupling as represented by 
the form factor in Eq. (\ref{formf}).

\begin{figure}[htbp]
\centerline{
\includegraphics[width=7.5cm,angle=90]{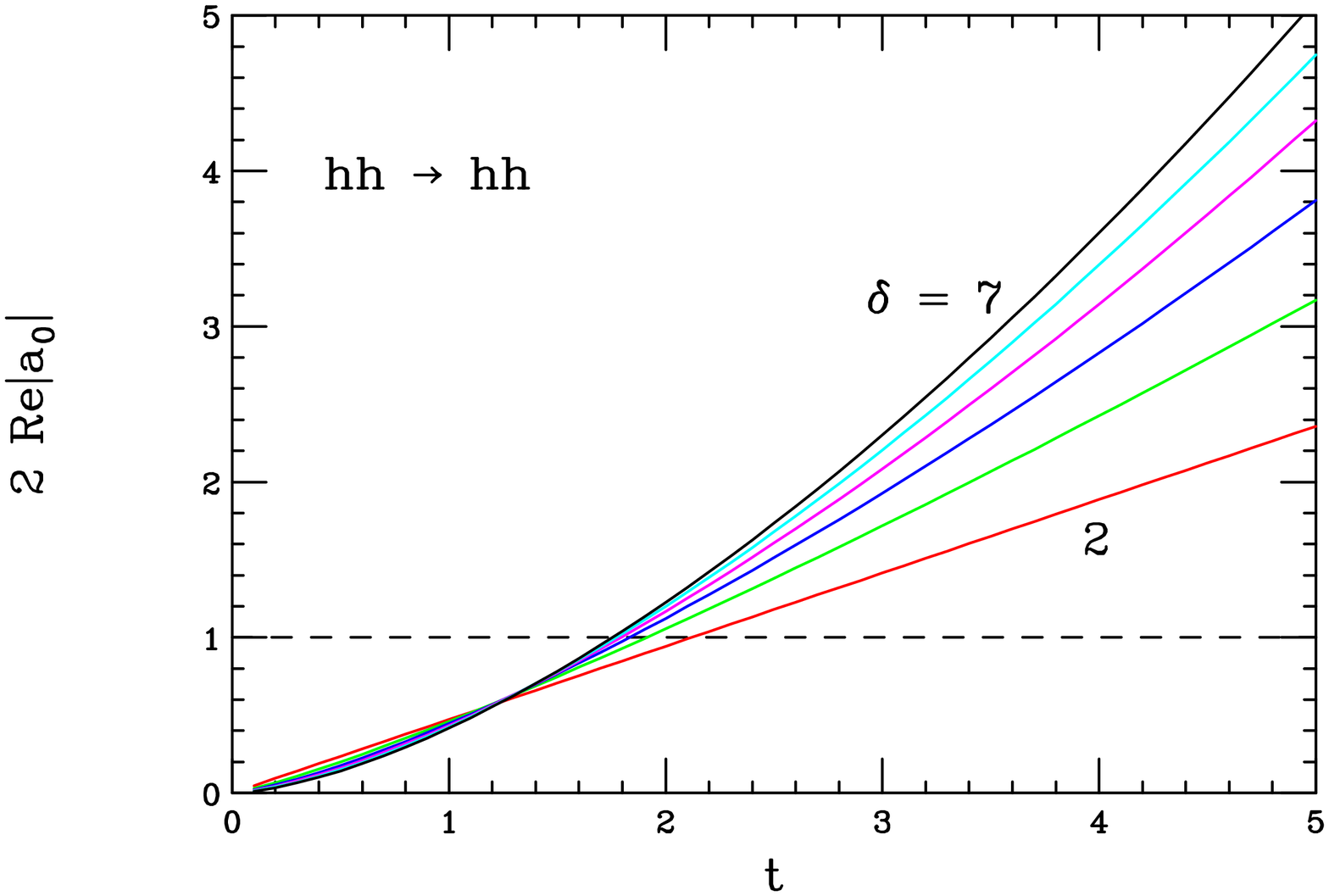}}
\vspace*{0.15cm}
\centerline{
\includegraphics[width=7.5cm,angle=90]{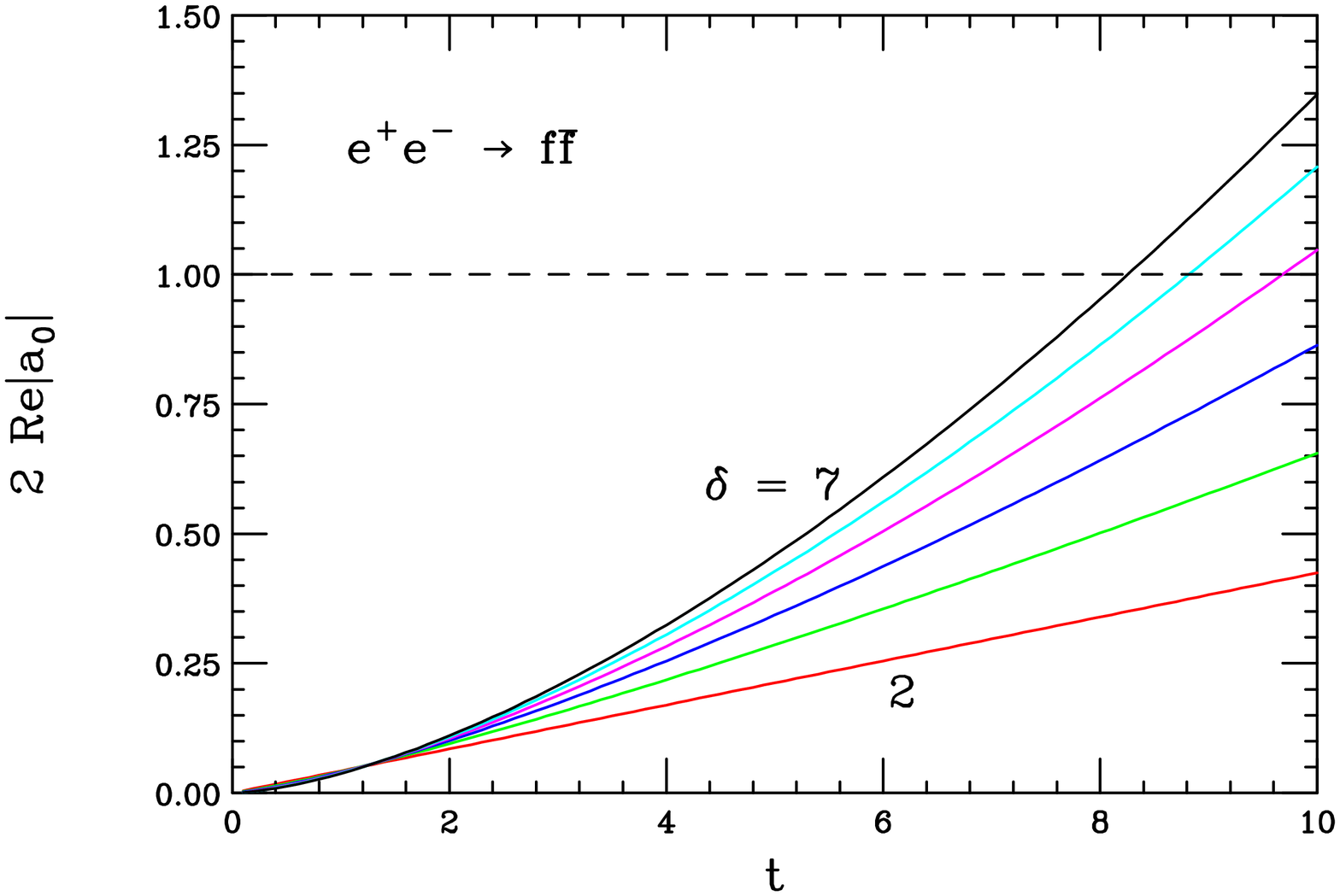}}
\vspace*{0.1cm}
\caption{The maximum value of the $J=0$ partial wave scattering amplitude as
$\sqrt s/\Lambda_H$ is varied
for (top panel) $hh\to hh$ and (bottom panel) $\epem\to f\bar f$  
as a function of $t$.  The curves correspond
to $\delta=2,3,4,5,6,7$ from bottom to top on the right-hand side.}
\label{unit}
\end{figure}

We first examine the effects of the form factor in unitarity considerations in 
high energy $2\to 2$ scattering.  In this model, the Kaluza-Klein (KK) 
tower of gravitons contribute to such processes via virtual multi-channel 
exchanges.  The operator for this transition takes the generic 
form \cite{me,grw}
\begin{equation}
 {\cal L} \simeq  i{4\over \Lambda_H^4}T^{\mu\nu}T_{\mu\nu} + h.c.\,,
\label{exchange}
\end{equation}
where the scale $\Lambda_H$ represents a naive cutoff that regulates the
integral over the sum of KK graviton 
propagators weighted by the density of KK states.  
$\Lambda_H$ is of order the fundamental scale $M$, with the exact relationship 
between the two being governed by the full UV theory.  At high
energies, the $2\to 2$ $s$-channel graviton exchange 
amplitude grows as, \eg, $s^2/\Lambda_H^4$, and thus
exhibits extremely poor behavior in the UV limit, violating perturbative
unitarity when $\sqrt s\gsim\Lambda_H$.  
Here, we explore whether the $\mu /tM$ term in the form
factor governing the running gravitational coupling 
can regulate this amplitude at high energies for some values of the
parameter $t$.  We first examine the case of $2\to 2$ Higgs boson
scattering, \ie, $hh\to hh$, as that is claimed \cite{He:1999qv} to be
the most sensitive process to the UV behavior of this theory.
Including the form factor as discussed in the previous section, setting
$\mu^2=s$, and computing the tree-level $J=0$ partial wave 
amplitude $a_0$ for this scattering process, 
we obtain the results displayed in Fig.~\ref{unit}. Here,
we show the maximum value that $2Re |a_0|$ obtains as the ratio 
$\sqrt s/\Lambda_H$ is varied, while holding $t$ and $\delta$, the number
of extra dimensions, fixed.  
A good test of unitarity in $2\to 2$
scattering is that the zeroth partial wave amplitude be bounded
by $|a_0|<1/2$ for all values of $s$.  We see that for all values of $\delta$ 
this condition is satisfied once the form factor is included, 
provided the parameter $t\lsim 2$.  As a second example, we also examine
the process $\epem\to f\bar f$, following the same procedure.  
The results for the maximum value of the
zeroth order partial wave amplitude for this process 
are shown in the bottom panel of
Fig.~\ref{unit}.  We see that in this case, the bounds from perturbative
unitarity are somewhat weaker, 
requiring only that $t\lsim 8$ or so.  Note that the
values of $t$ which allow for good UV 
behavior of this theory agree with
the more theoretical expectations discussed in the previous section, which 
predicted $t$ to be of order unity.  Following these guidelines, 
we will take $t\le 2$ in our ensuing calculations.

\begin{figure}[htbp]
\centerline{
\includegraphics[width=7.5cm,angle=90]{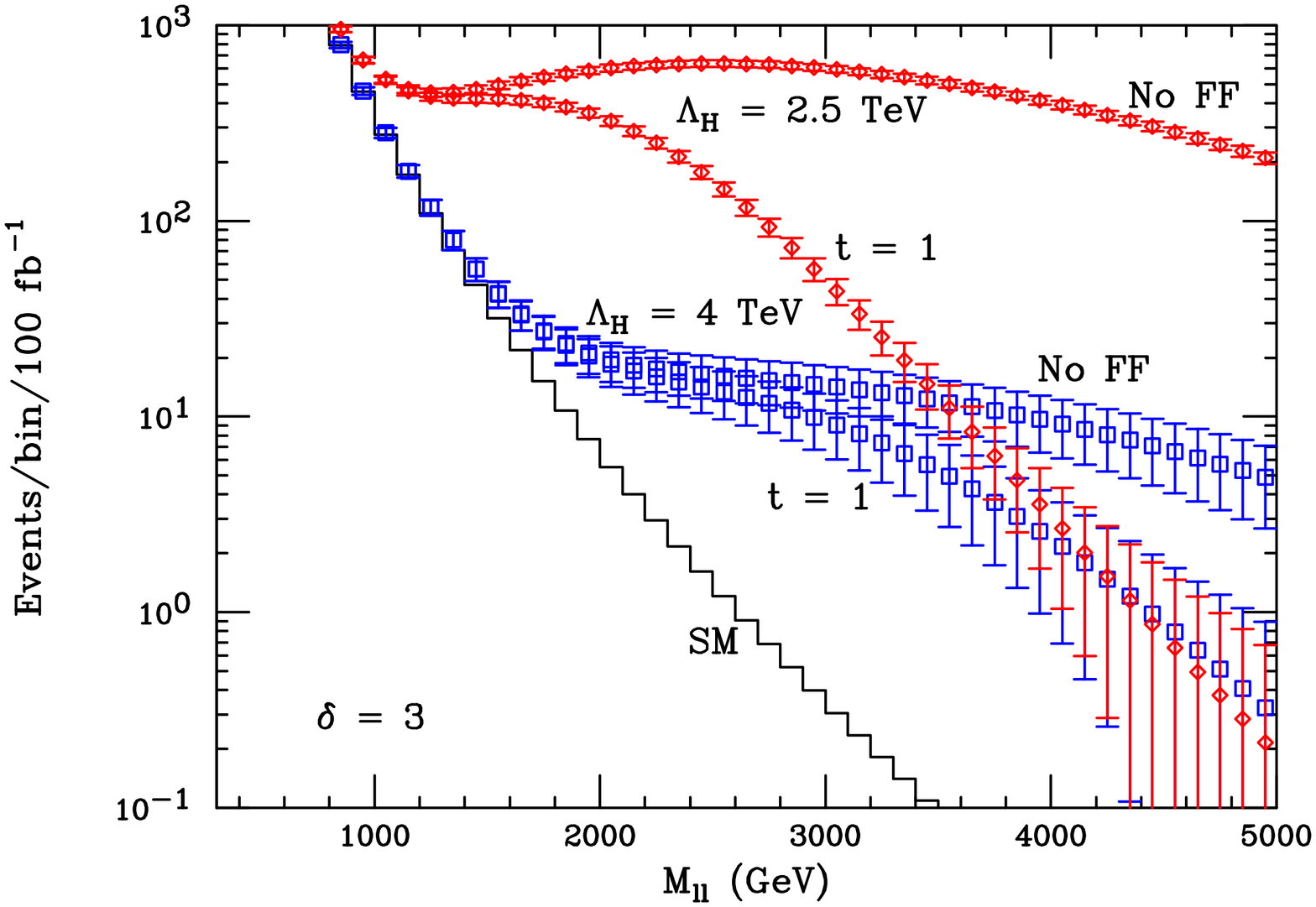}}
\vspace*{0.1cm}
\caption{The event rate per bin with 100 \infb\ of integrated luminosity 
for Drell-Yan production at the LHC as a function of the
lepton pair invariant mass with $\delta=3$.  
The red (blue) curves correspond to scale $\Lambda_H
=2.5~(4.0)$ TeV.  In each case, the top curve is the result in the
ADD model without the form factor \cite{me}, and the bottom curve includes
the form factor with $t=1$.  The error bars represent the statistical
errors.  The black histogram corresponds to the Standard Model event rate.}
\label{lhccomp}
\end{figure}

We now turn to the conventional collider signatures of this scenario.  
The first class of processes that we consider is the contribution of 
virtual KK graviton exchange in
Drell-Yan production, $pp\to\ell^+\ell^- +X$, at the LHC. 
This contribution proceeds through the operator given in Eq.~\ref{exchange}, 
and involves $q\bar q$ and $gg$ initial parton states.  The same
cutoff scale $\Lambda_H$ is employed and when including the form factor
parameterizing running gravitational couplings,
the scale $\mu$ in the form factor
is set to $\sqrt{\hat s}$ as discussed above.  The unmodified
graviton exchange
amplitude behaves as $\sim \hat s^2/\Lambda_H^4$ and we expect the
form factor to modify the production cross section at high energies.
Figure \ref{lhccomp} shows the number of events for 100 \infb\ of integrated
luminosity as a
function of the invariant mass of the final state lepton pair, with and without
the form factor, taking $t=1$ and $\delta=3$.  We see that the form
factor has a dramatic effect on the production cross section and results in a
sharp reduction of the event rate at high invariant masses.

\begin{figure}[htbp]
\centerline{
\includegraphics[width=6cm,angle=90]{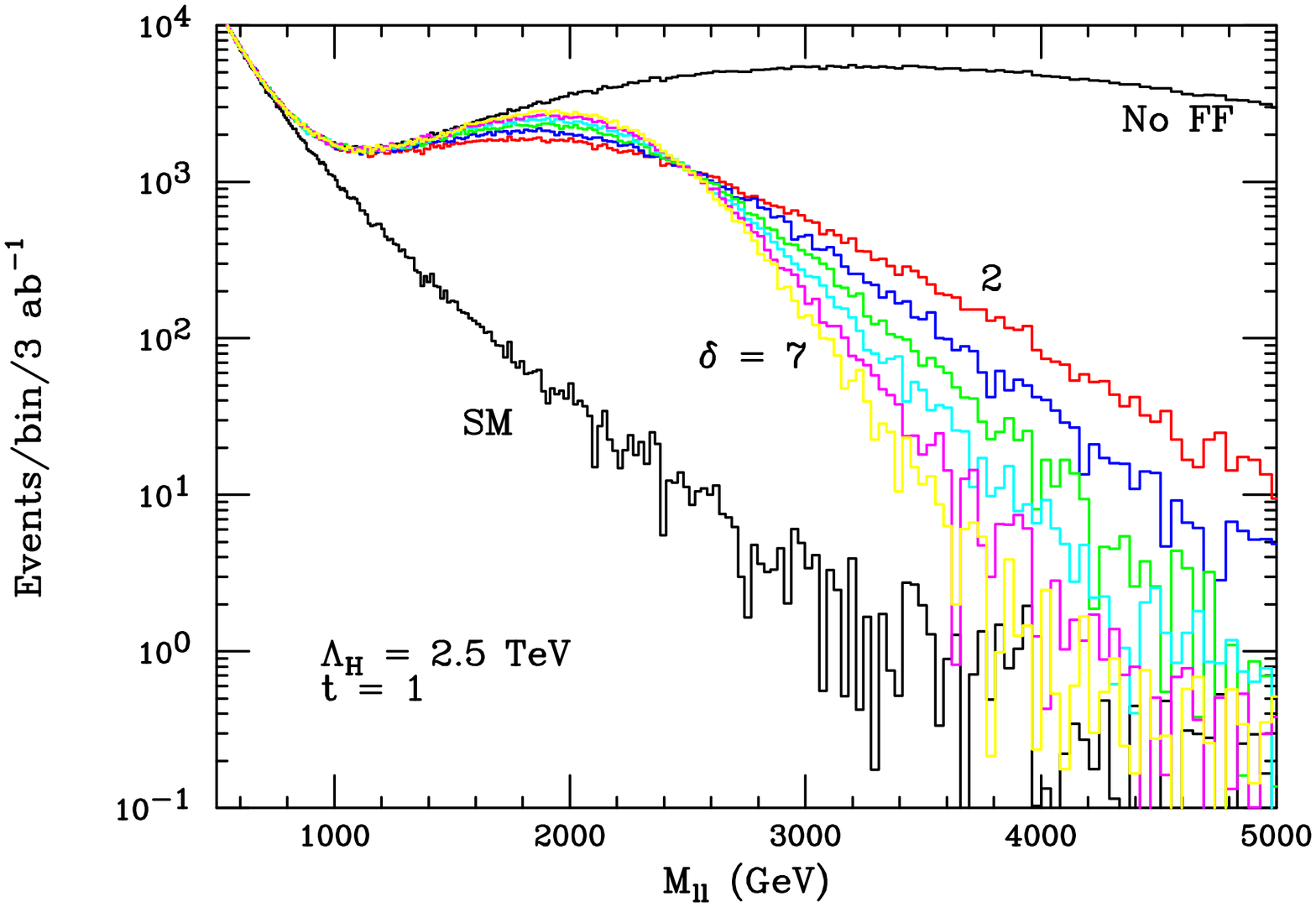}
\hspace*{5mm}
\includegraphics[width=6cm,angle=90]{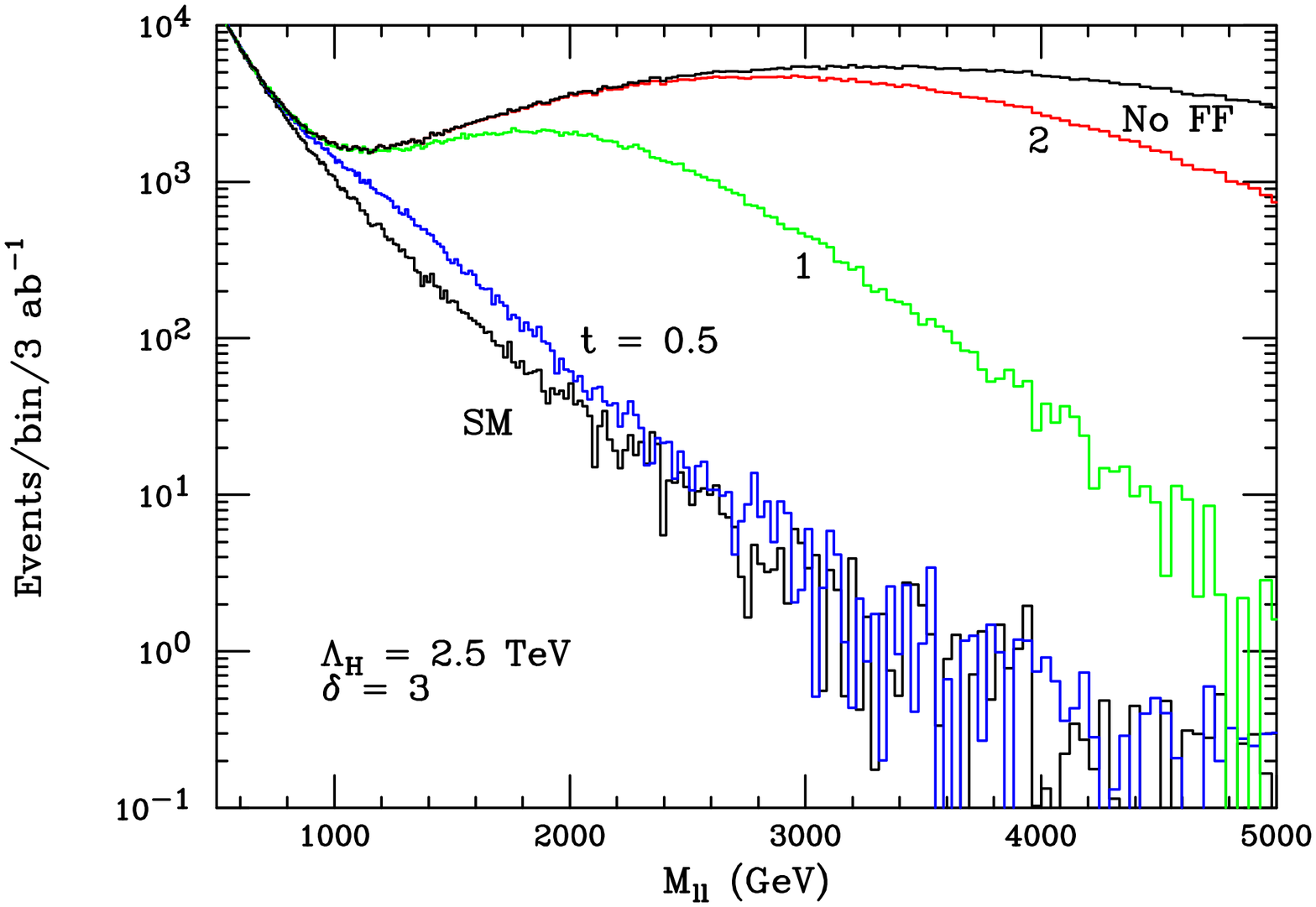}}
\vspace*{0.15cm}
\centerline{
\includegraphics[width=6cm,angle=90]{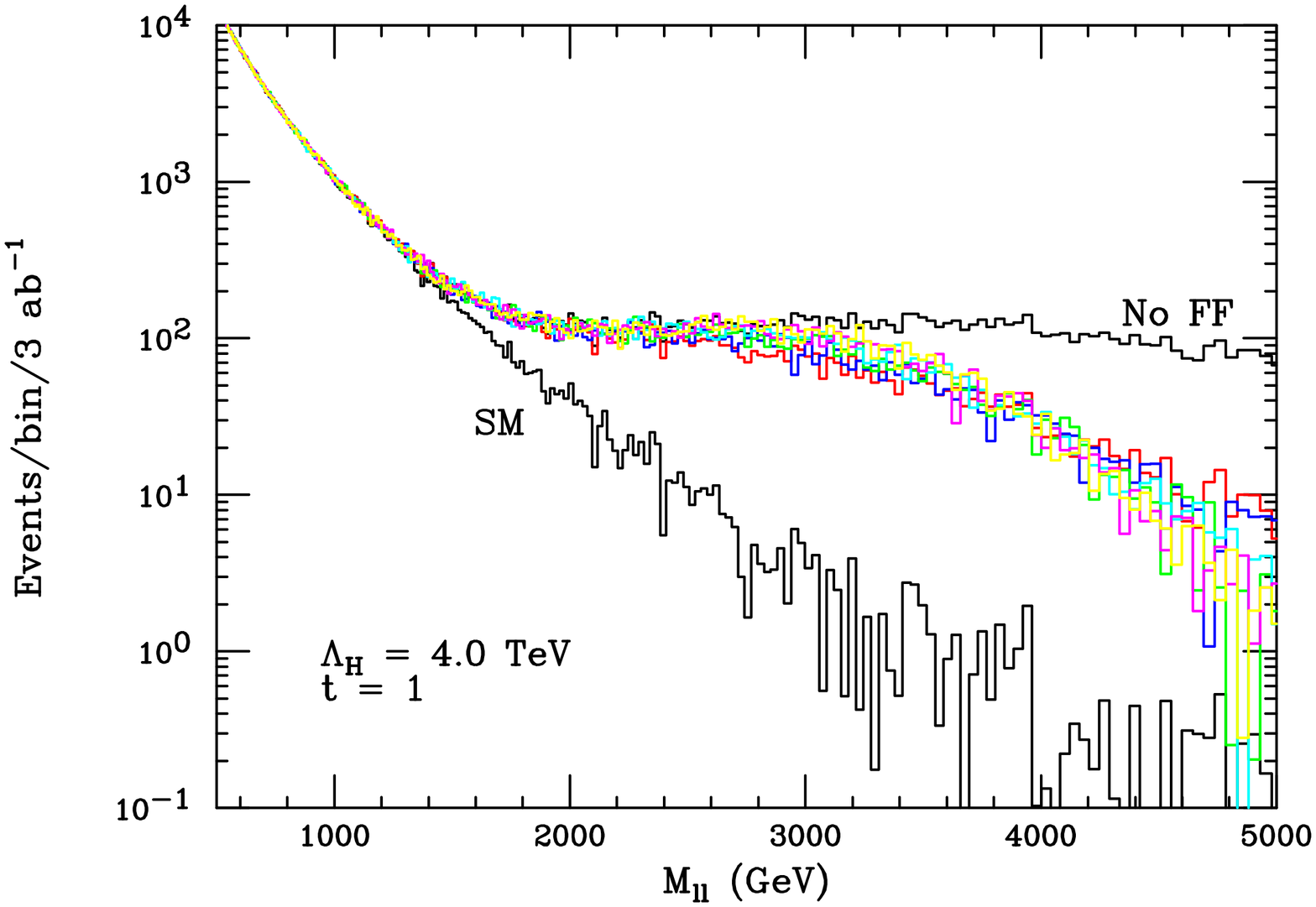}
\hspace*{5mm}
\includegraphics[width=6cm,angle=90]{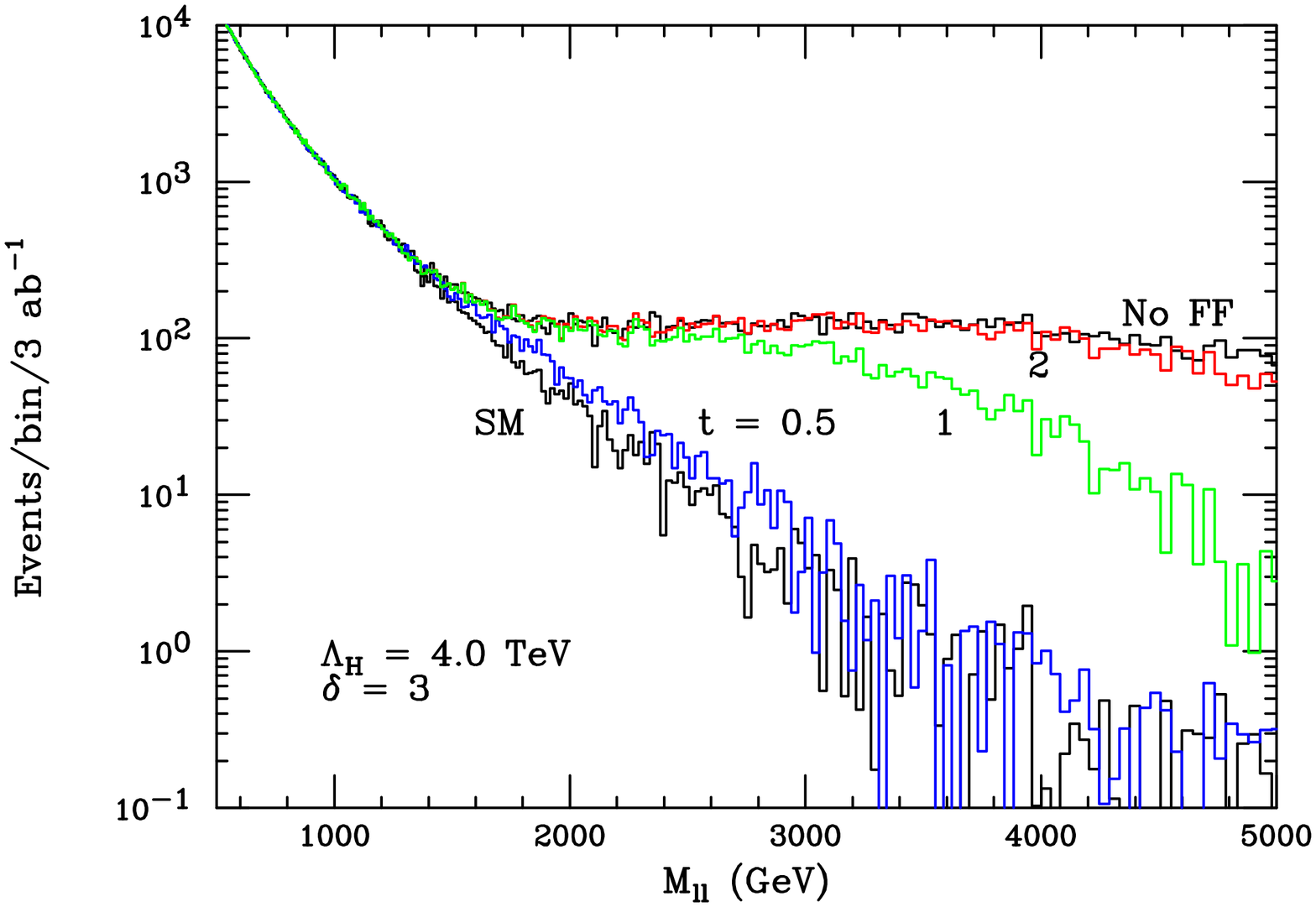}}
\vspace*{0.1cm}
\caption{The event rate per bin with 3 ab$^{-1}$ of integrated luminosity 
for Drell-Yan
production at the LHC as a function of the lepton pair invariant
mass taking the scale $\Lambda_H=2.5$ TeV (top panels) and
$\Lambda_H=4.0$ TeV (bottom panels).  In the left panels, $t=1$ and
$\delta=2,3,4,5,6,7$ from top to bottom on the right-hand side as
labeled.   In the right panels, $\delta=3$
and the red, green, blue curves correspond to $t=2,1,0.5$.  In all
panels,  the bottom black histogram corresponds to the Standard Model
result, and the top black histogram is the conventional ADD result
\cite{me} without the form factor.}
\label{lhcexch}
\end{figure}

We examine these effects in more detail in Fig.~\ref{lhcexch}.  Here, we
take $\Lambda_H=2.5$ and 4.0 TeV and separately study the variation in the
Drell-Yan production cross section due to $t$
and $\delta$.  In this computation, the size of the statistical errors
in the event rate are
shown by the size of the fluctuations in the binned rate. 
Holding the parameter $t$ fixed at $t=1$ and varying the number of
extra dimensions, we see that the effect of the form factor is more
pronounced and separated for the various values of $\delta$ at 
high invariant masses.  In the formalism employed here, the cross section
for graviton exchange is insensitive to the number of extra dimensions.
However, note that here, due to the form factor, the production rate is larger
with increasing (decreasing) values of $\delta$ when the lepton pair
invariant mass is less than (greater than) $t\Lambda_H$.
Holding $\delta$ fixed, we see that decreasing the value of 
$t$ sharply increases the effects of the running gravitational coupling,
as we would expect.  In fact, for $t=1/2$, the contribution of virtual
KK graviton exchange is damped to the point where the event rate lies
not far above that of the Standard Model, particularly at large invariant
masses.  However, for $t=2$, the change
in the event rate is only minor
in comparison to that of the standard ADD model. 
Hence, as $t$ is varied over its
theoretically expected range, the size of the form factor effect on 
virtual graviton exchange in models with large extra dimensions 
differs greatly.

\begin{figure}[t]
\centerline{
\includegraphics[width=7.5cm,angle=90]{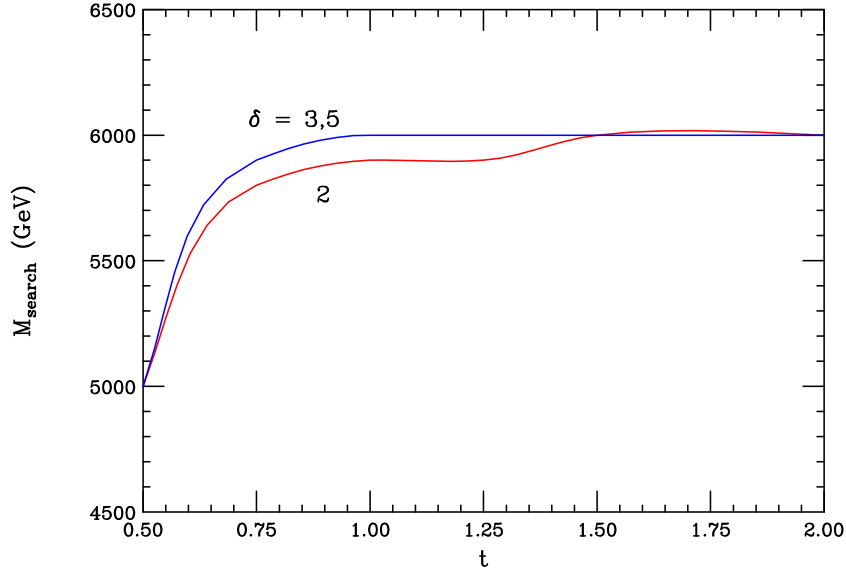}}
\vspace*{0.1cm}
\caption{The 95\% C.L. search reach for the fundamental scale $M$ in
Drell Yan production at the LHC as a function of the parameter $t$.  
The red curve
corresponds to $\delta = 2$, while the $\delta=3$ and 5 curves lie on
top of each other and are represented by the blue curve.}
\label{dylim}
\end{figure}

Given the striking effect of a running gravitational coupling in the
Drell-Yan invariant mass distribution at the LHC, 
we now address the question of whether the search reach for the 
scale $\Lambda_H$ is modified in the presence of the form factor.
The 95\% C.L. search reach for the cutoff scale $\Lambda_H$ in this process 
is presented in Fig.~\ref{dylim}
as a function of $t$ for various values of $\delta$.  We see that
for $t\gsim 1$, the search reach is unaffected by the presence of the
form factor, since the cross section is independent of $\delta$ for large
values of $t$.  For $t\lsim 1$, the reach degrades with decreasing $t$,
but not substantially. For example, the reach in $\Lambda_H$ decreases by only
$\simeq 1$ TeV when $t$ takes on the value of 0.5.  
This is due to the large statistics
available at the LHC for lower values of the lepton pair invariant
mass, which coincides with the region where the running coupling has
the smallest effect.

We now consider the class of collider processes that involve the
real emission of KK graviton states at the LHC, \ie, the scattering process
$pp\to jet +G_n$, where $G_n$ represents a state in the
graviton KK tower \cite{grw,mp}.  
The produced graviton behaves as if it were a massive,
non-interacting, stable particle and once the KK states are summed, it yields 
a distribution of missing energy.  
The specific process kinematics regulate this reaction and
there is no need to introduce a cutoff.
This jet plus missing energy signature arises from the three 
sub-processes $gg,q\bar q\to gG_n$ and $gq\to qG_n$, and results in a
$\delta-$dependent reach directly on the fundamental scale $M$. 
The search reach for this reaction, together 
with the SM backgrounds, have been well studied by the authors
in \cite{Vacavant:2001sd}; 
they find, for example, that if $\delta=2~(3,~4,~5,~6)$ the maximum reach 
on $M$ at the LHC is 9.1 (7.0, 6.0, 5.5, 5.2) TeV, respectively, for an
integrated luminosity of 100 \infb. 
As discussed in the previous Section, a study of the effects
of the running gravitation coupling is more complicated for this reaction
as the various sub-processes do not take place at a fixed center of mass 
energy and several different scales are present.  
This introduces an ambiguity in the choice of the scale $\mu$
in the form factor.  Here, as examples, 
we examine two possibilities:  $\mu = E_{jet}$
or $p_{T,jet}$.

\begin{figure}[htbp]
\centerline{
\includegraphics[width=6cm,angle=90]{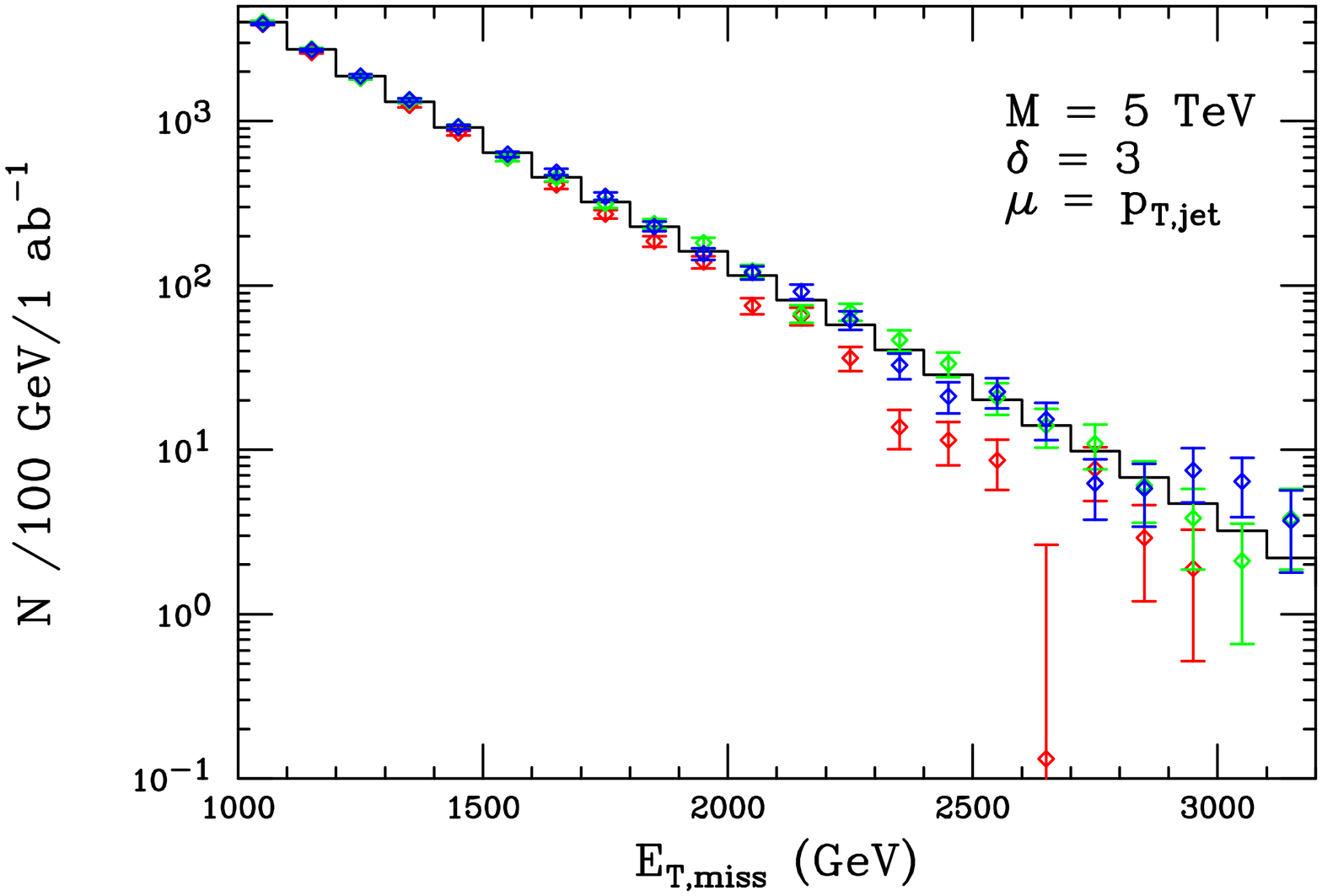}
\hspace*{5mm}
\includegraphics[width=6cm,angle=90]{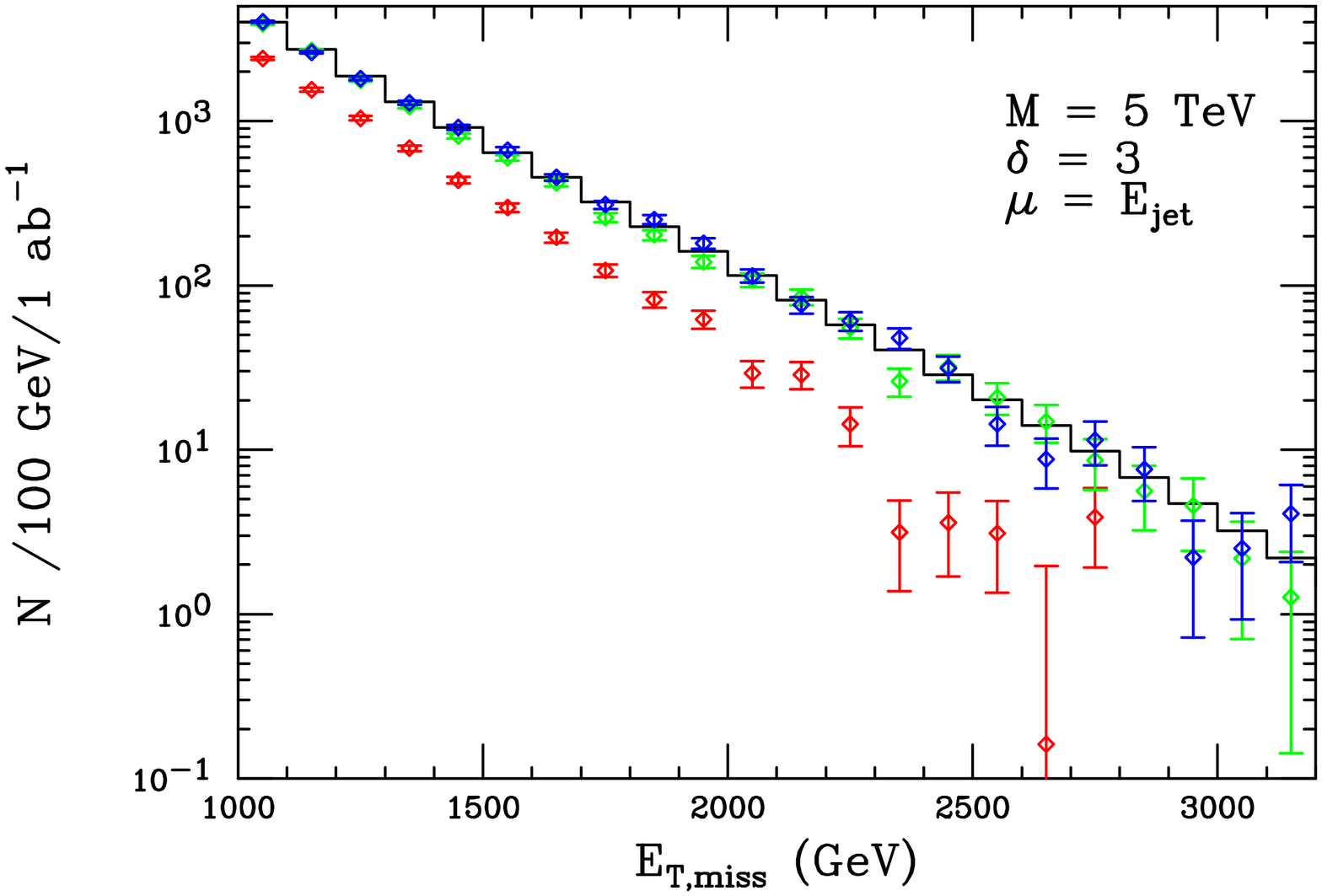}}
\vspace*{0.15cm}
\centerline{
\includegraphics[width=6cm,angle=90]{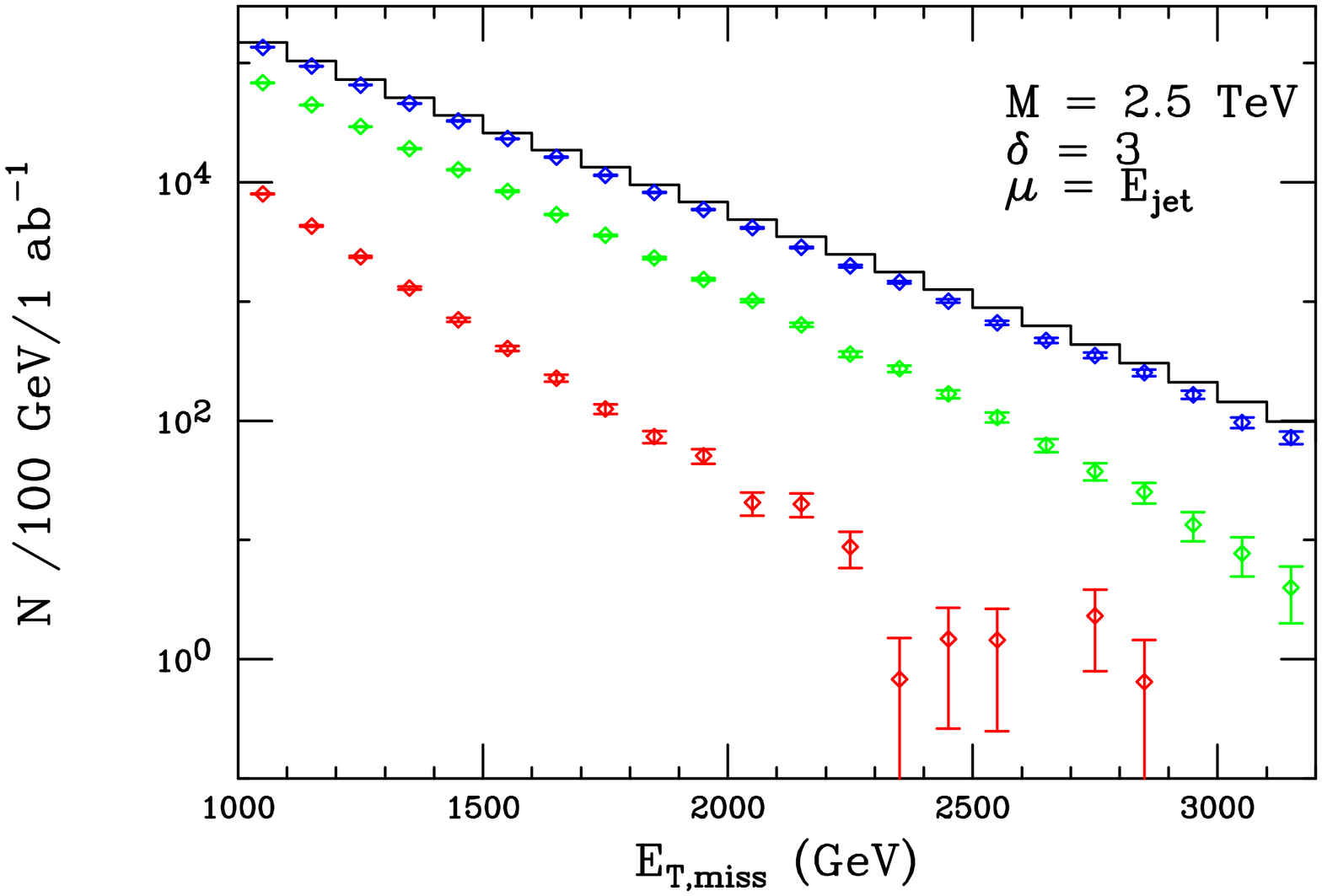}
\hspace*{5mm}
\includegraphics[width=6cm,angle=90]{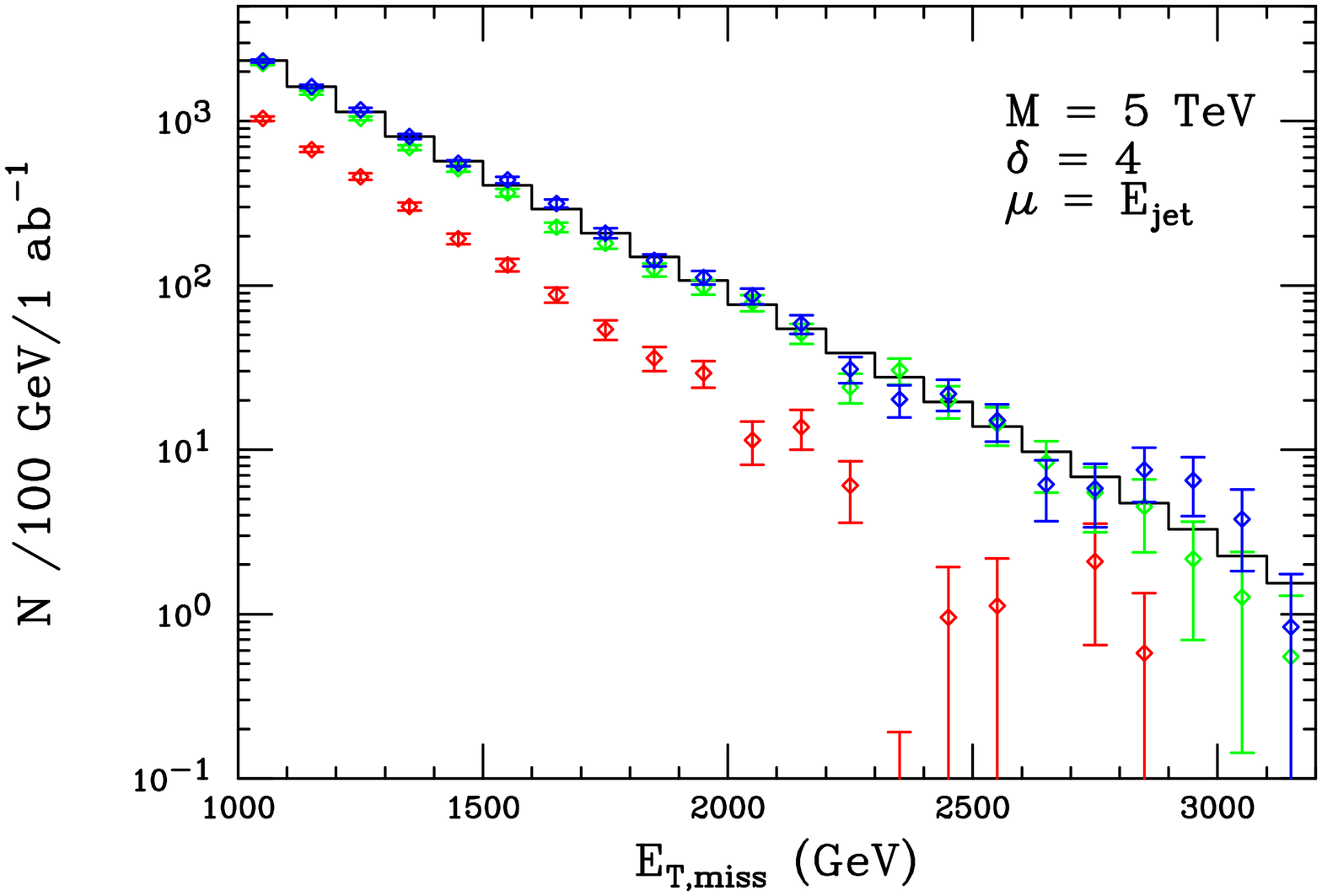}}
\vspace*{0.1cm}
\caption{Missing transverse energy distribution for the signal process
$pp\to jet +\not E_T$ assuming 1 ab$^{-1}$ of
integrated luminosity at the LHC.  The standard ADD result is given
by the black histogram, while the blue, green, and red data points correspond
to the inclusion of the form factor with $t=2,~1,~0.5$, respectively.
The other parameters are as labeled.  The errors bars represent the
statistical errors.}
\label{lhcemit1}
\end{figure}

Figure~\ref{lhcemit1} displays the missing energy distribution for
the signal at the LHC for this process with $t=0.5,~1,~2$.  The top 
two panels compare the choices $\mu=E_{jet}$ and $p_{T,jet}$ for fixed
values of $M$ and $\delta$.  We see that the choice of $\mu=p_{T,jet}$
yields a much smaller deviation from the conventional result without the
form factor than does the
case of $\mu=E_{jet}$. In both cases, sizeable modifications of
the distribution only
occur when $t$ takes on the value of 0.5, and become more pronounced 
at large values of missing $E_T$.  This is to be expected since in general
$p_{T,jet}<E_{jet}$ and the effect of the form factor grows
as the scale $\mu$ increases in magnitude.
The bottom two panels explore the
form factor effects when $M$ and $\delta$ are varied, taking $\mu=E_{jet}$.
Here, we see that this effect is amplified for
lower values of $M$ and larger values of $\delta$, yielding a significantly
smaller missing $E_T$ distribution than in the standard case.

\begin{figure}[htbp]
\centerline{
\includegraphics[width=7cm,angle=90]{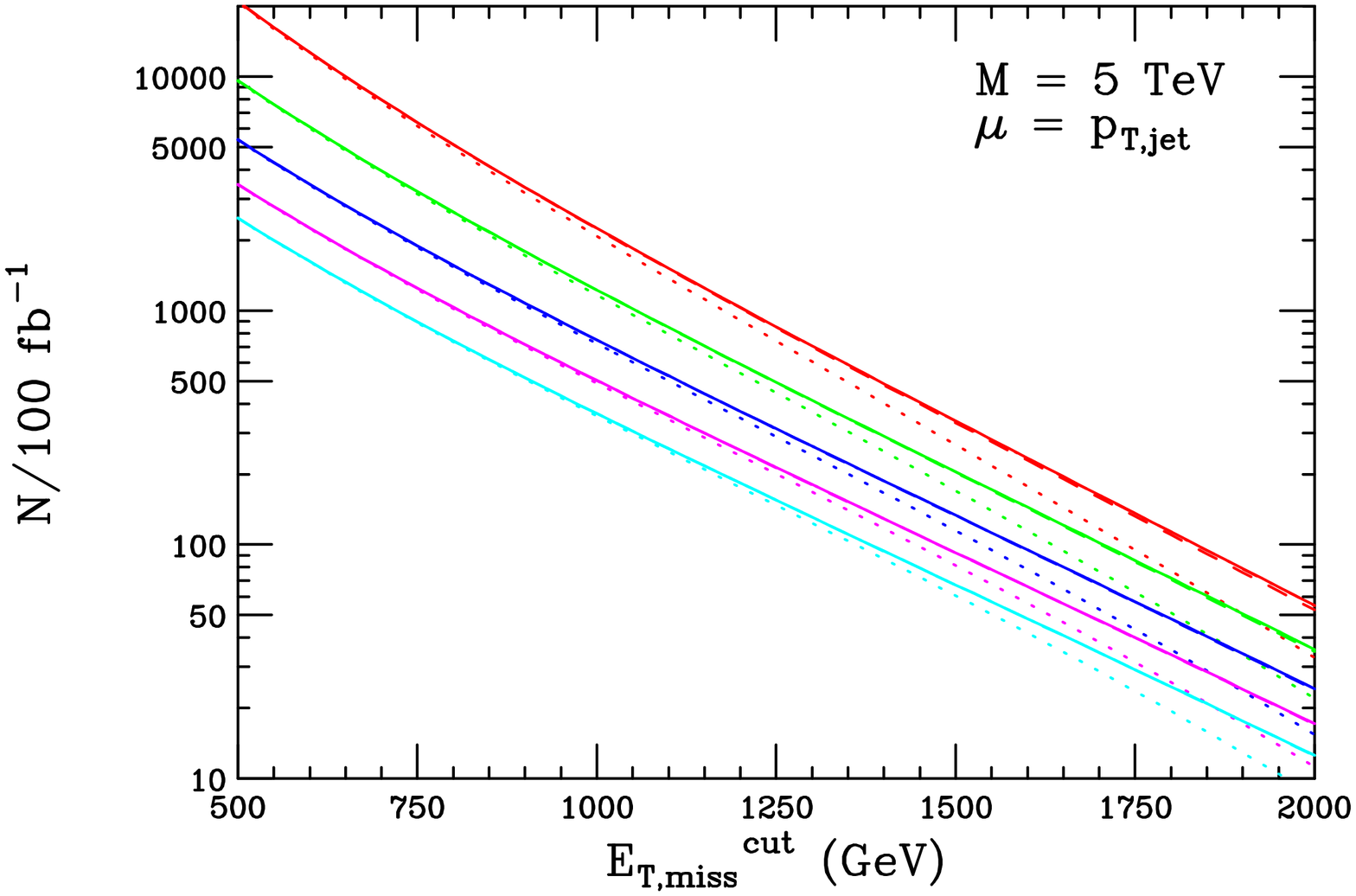}}
\vspace*{0.15cm}
\centerline{
\includegraphics[width=7cm,angle=90]{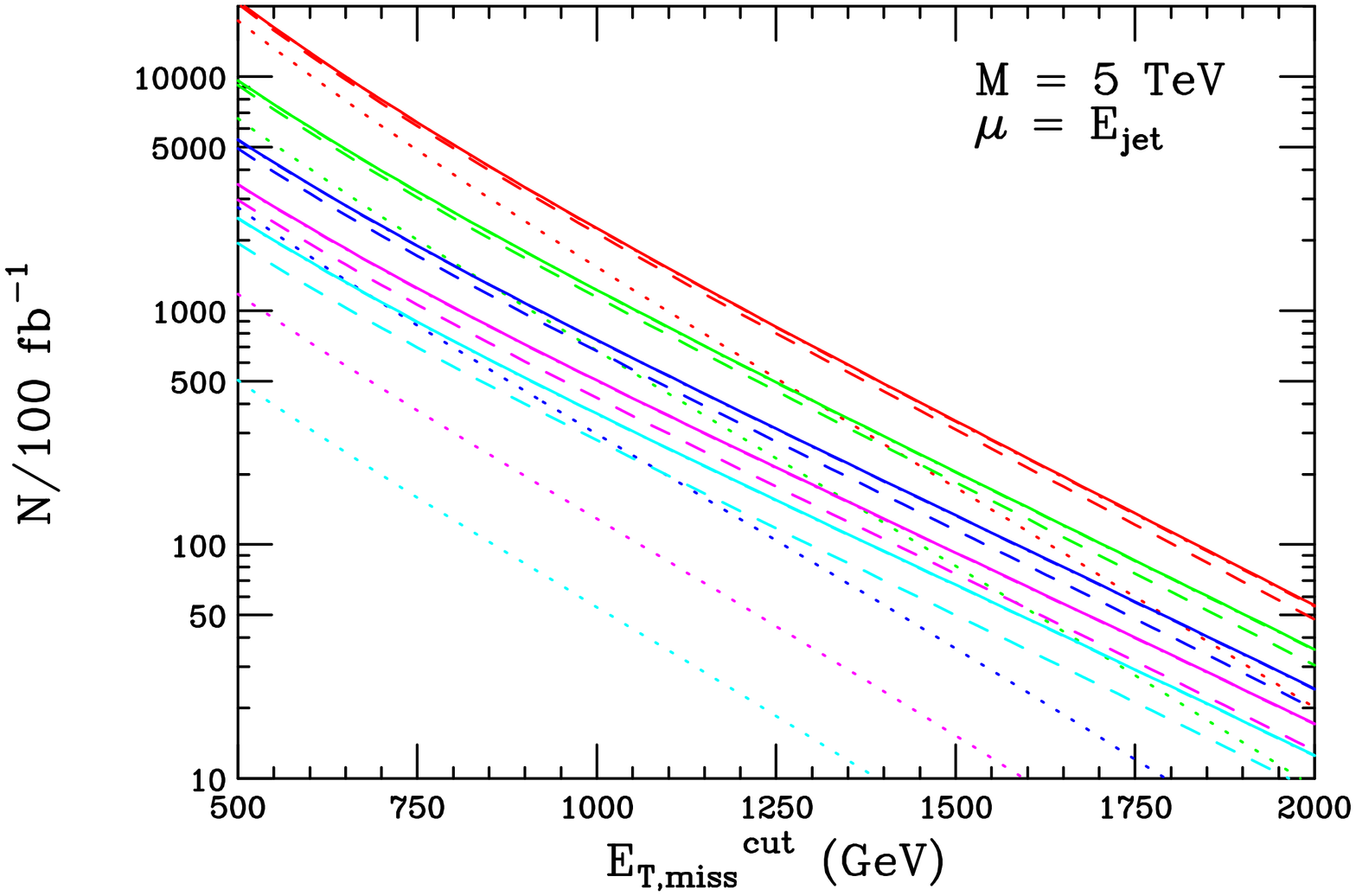}}
\vspace*{0.1cm}
\caption{The excess 
signal event rate for $pp\to jet +G_n$ with 100 fb$^{-1}$ of
integrated luminosity at the LHC as a function of a cut on missing $E_T$.
$M=5$ TeV, $\delta=2,3,4,5,6$ from top to bottom, and 
$\mu=p_{T,jet},~(E_{jet})$ in the top (bottom) panel.  The solid curves
correspond to the conventional ADD result and the (invisible) dash-dotted,
dashed, and dotted curves are for $t=2,~1,~0.5$, respectively.}
\label{lhcemit2}
\end{figure}

\begin{figure}[htbp]
\centerline{
\includegraphics[width=7cm,angle=90]{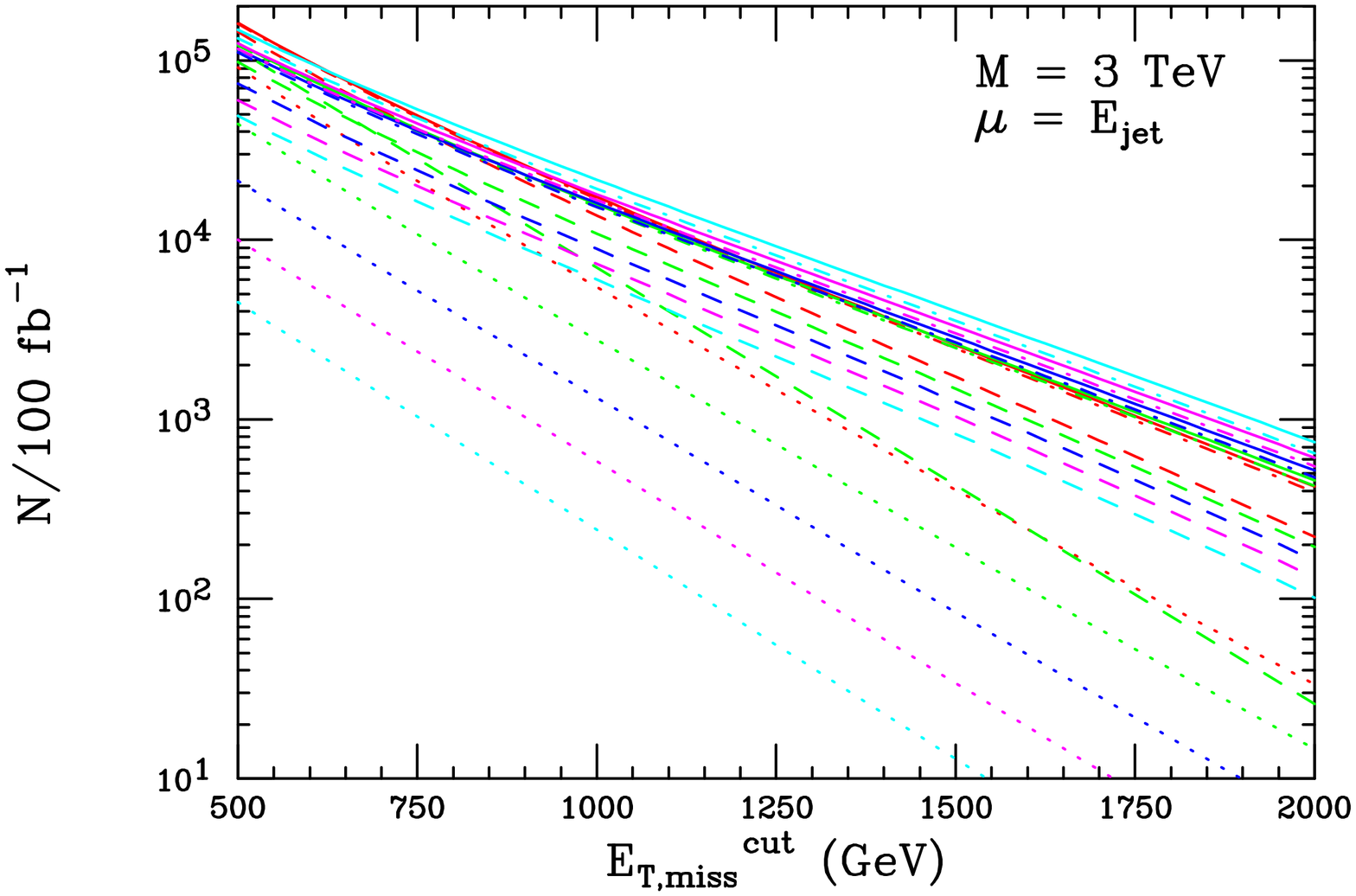}}
\vspace*{0.15cm}
\centerline{
\includegraphics[width=7cm,angle=90]{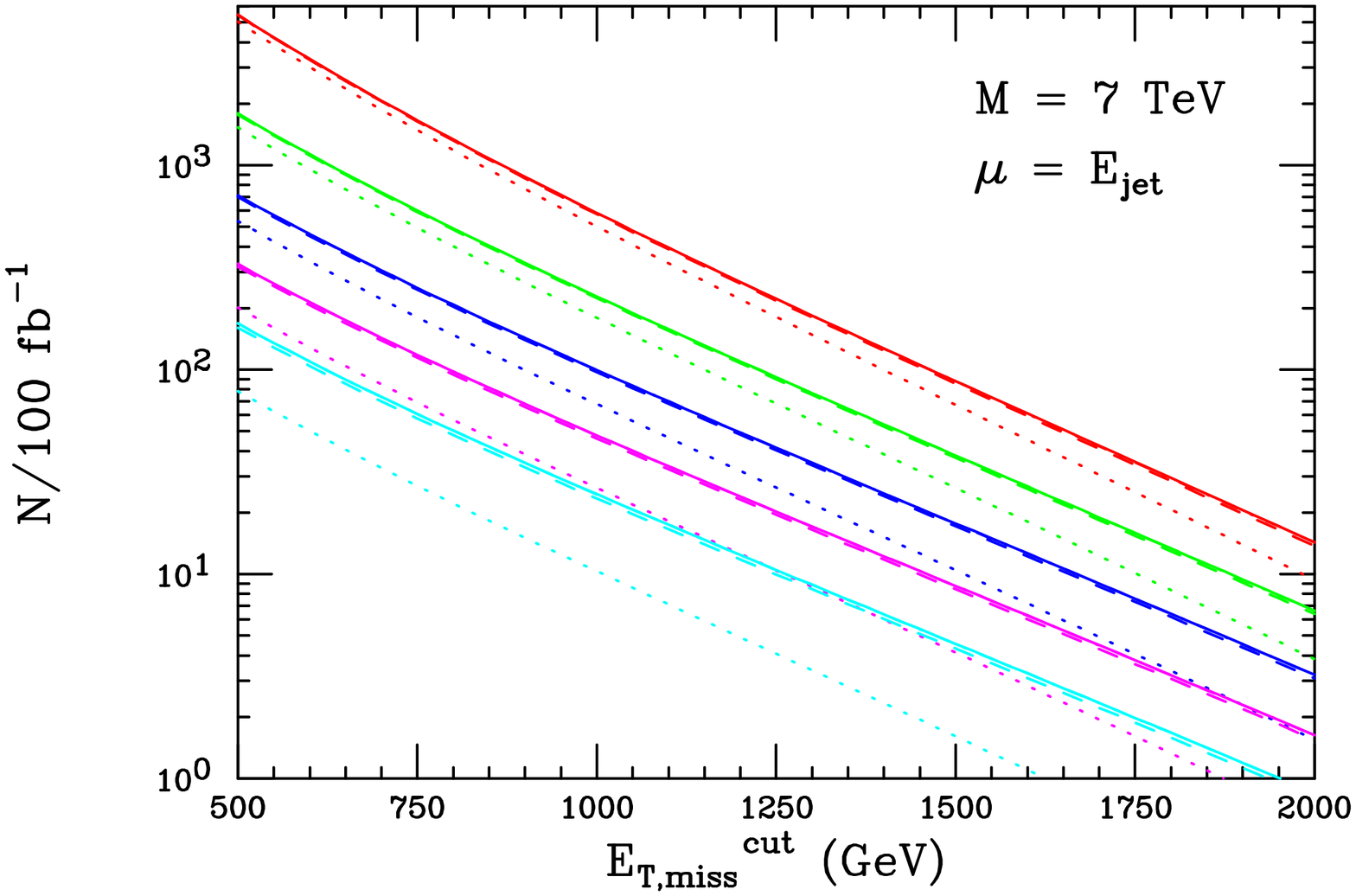}}
\vspace*{0.1cm}
\caption{The excess 
signal event rate for $pp\to jet +G_n$ with 100 fb$^{-1}$ of
integrated luminosity at the LHC as a function of a cut on missing $E_T$.
$\delta=2,3,4,5,6$ from top to bottom, 
$\mu=E_{jet}$, and $M=3,~(7)$ TeV in the top (bottom) panel.  The solid curves
correspond to the conventional ADD result and the (invisible) dash-dotted,
dashed, and dotted curves are for $t=2,~1,~0.5$, respectively}
\label{lhcemit3}
\end{figure}

Next we examine the total event rate for the signal 
above a cut on missing $E_T$, as a
function of that cut.  Note that strong cuts are required on missing $E_T$
in order to suppress the Standard Model background.  
Figure~\ref{lhcemit2} compares the choices
$\mu=E_{jet}$ and $p_{T,jet}$ for $M=5$ TeV for various values of the
parameter $t$ and the number of extra dimensions $\delta$.  Again, 
larger deviations from the conventional ADD result are
obtained in the case $\mu=E_{jet}$ and the effect of the form factor
increases as $t$ becomes smaller.  The case of $t=2$ is essentially 
indistinguishable
from the standard ADD result.   Figure~\ref{lhcemit3} shows the
consequences for different values of the fundamental scale $M$.  Here,
we see that the effects of a running gravitational coupling are quite
significant for smaller values of $M$, as would be expected.

We now study the modification to the search reach for large extra
dimensions in this channel.
Using the same search criteria as in \cite{Vacavant:2001sd}, we examine the
case which yields the largest deviation from the conventional ADD result,
\ie, we assume $t=0.5$ and take $\mu=E_{jet}$, and find that
the search reaches are reduced to 
9.0 (6.6, 5.3, 4.2, 3.0) TeV, for $\delta=2~(3,~4,~5,~6)$, respectively.
Note that the search reach degradation in comparison to the conventional ADD
case increases for larger numbers of extra dimensions. 
Choosing $\mu=p_{T,jet}$ instead, and assuming the 
same value of $t$, we find that there are essentially 
{\it no} modifications in the 
search reach from the conventional results. For either choice of $\mu$, 
taking $t\geq 1$, yields no reduction in the ADD 
search reach in this channel. Thus the ability to see the jet plus 
missing energy signature of large extra dimensions 
remains rather robust when form 
factor effects are included, as long as the parameter $t$ is not too small.

Next, we examine the signatures of a running gravitational coupling at
the ILC.  The basic processes that are relevant for the ADD scenario
are virtual KK graviton exchange and the direct production of KK gravitons 
via graviton emission as discussed above for the LHC.

\begin{figure}[htbp]
\centerline{
\includegraphics[width=7.0cm,angle=90]{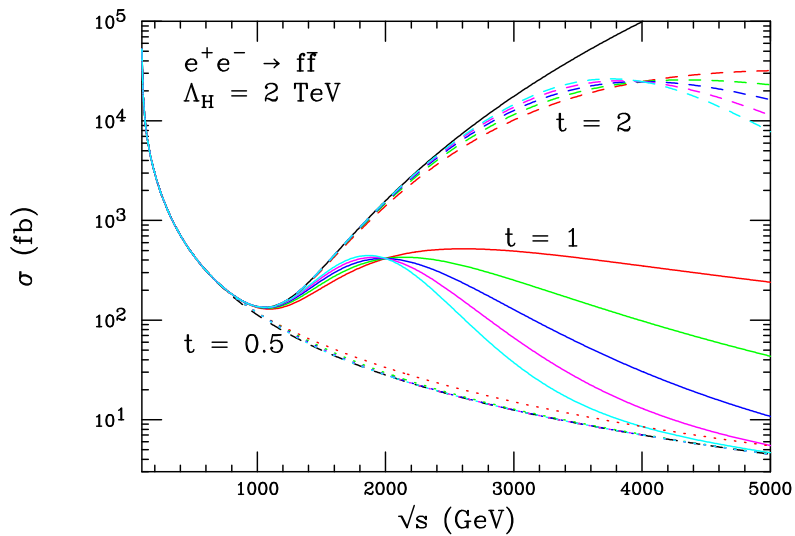}}
\vspace*{0.1cm}
\caption{The cross section for $\epem\to f\bar f$ as a function of
$\sqrt s$ with $\Lambda_H=2$ TeV.  The solid, dashed, and dotted sets of
curves correspond to the parameter $t=2,~1,~0,5$, respectively, and
$\delta=2,~3,~,4~,5,~6$ correspond to the curves in each set from top
to bottom on the right-hand side.  The solid black curves at the top
(bottom) represent the conventional ADD (Standard Model) result for this
channel.}
\label{ilcexch}
\end{figure}

We first consider the case of graviton exchange in the reaction
$\epem\to f\bar f$.  At the $\sqrt s=500$ GeV ILC, the search reach
for the cutoff $\Lambda_H$ in the conventional ADD model is
approximately 5 TeV \cite{me,teslatdr}, independent of the value of 
$\delta$, assuming an integrated luminosity of 500 $fb^{-1}$ 
and $80\%$ electron beam polarization.  Since $\sqrt s$ is 
fixed in this channel, the implementation of the form factor 
is straightforward and proceeds as discussed above.  
The modifications in the cross section as a
function of $\sqrt s$ in the
presence of the form factor are illustrated in Fig.~\ref{ilcexch}, for
various values of $\delta$ and the parameter $t$, taking
$\Lambda_H=2$ TeV.  We see that the behavior of the cross section is
similar to that of the invariant mass distribution for Drell-Yan 
production at the LHC; the effects are more pronounced at small values
of $t$ and track the Standard Model result for $t=0.5$.  Note that again the
dependence on $\delta$ reverses as the threshold $\sqrt s=t\Lambda_H$
is passed.  At large enough values
of $\sqrt s$, the cross section for $t=2$ starts to turn over,
displaying the onset of unitarity as discussed above.  For this value
of $\Lambda_H$, a higher center-of-mass energy is clearly beneficial in order
to detect the presence of the form factor.

Since the implementation of the form factor 
is straightforward, the modification to the search reach can be
derived analytically.  We find (denoting the search reach with
running coupling as  $\Lambda_H^{FF}$)
\begin{equation}
\Lambda_H^4 =(\Lambda_H^{FF})^4+(\Lambda_H^{FF})^{-\delta+2}
\Big ({{\sqrt s}\over {t}} \Big)^{\delta+2}\,,
\end{equation}
which can be easily solved numerically. If the parameter $t \geq 1$, we find
that $\Lambda_H^{FF}$ does not differ from $\Lambda_H$ by 
more than $0.5\%$ for any value of $\delta$. If, however, 
$t=0.5$ and $\delta \geq 5$ there is a reasonable 
search reach degradation; in particular, we find, for $\delta=5~(6,~7)$ 
that $\Lambda_H^{FF}=4.77~(4.20,~3.44)$ TeV.

\begin{figure}[htbp]
\centerline{
\includegraphics[width=7.0cm,angle=90]{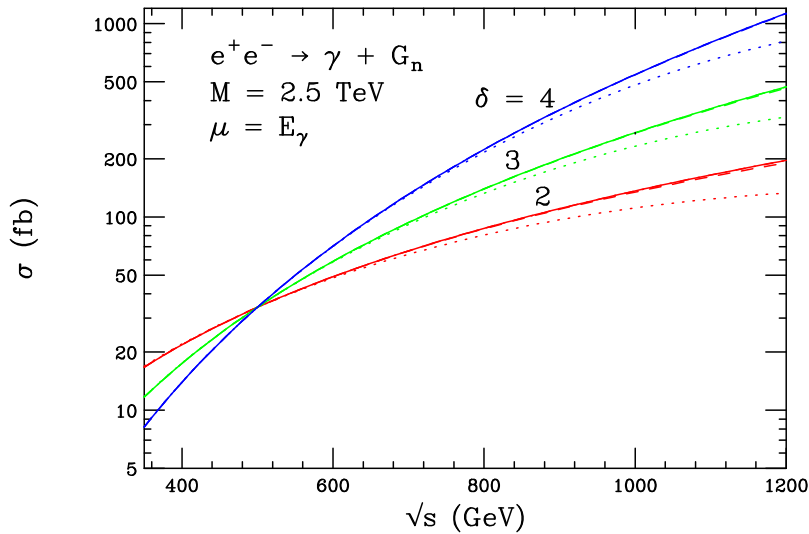}}
\vspace*{0.1cm}
\caption{The cross section for $\epem\to\gamma+\not E_T$ from KK graviton
emission as a function of
$\sqrt s$, taking $\mu=E_\gamma$.  All curves are normalized to the value
of the cross section at
$\sqrt s=500$ GeV with $M=2.5$ TeV and $\delta=2$.  
The (invisible) dash-dotted, dashed, and dotted curves 
correspond to
$t=2,~1,~0.5$, while the blue, green, and red sets of curves represent
$\delta=4,~3,~2$.  The solid curves represent the conventional ADD
cross section.}
\label{ilcemit}
\end{figure}

The last channel for us to consider is graviton emission at the ILC,
which proceeds via the reaction  $e^+e^-\to \gamma+G_n$.  The effect
of running gravitational couplings on the missing energy cross section 
are shown in
Fig.~\ref{ilcemit} as a function of $\sqrt s$.  In this figure, we
take $\mu=E_{\gamma}$ as the analogous choice was found to lead to the largest
deviations at the LHC.  We see that deviations from the conventional
result are only observable for $t=0.5$, which somewhat lowers the cross section
at larger values of $\sqrt s$.  Since this process is typically employed
as a means to determine the number of extra dimensions, the existence
of a form factor would in principle interfere with this determination.

An analytical analysis can also be applied for this process since the
center-of-mass energy is fixed.
In the usual 
case, the rate for graviton emission scales as $M^{-{(2+\delta)}}$ 
and one can obtain the reach for any given value of 
$\delta$ in a straightforward manner \cite{grw,mp}. 
Equating the signal rate that yields the search limit in the conventional 
ADD case to the rate in the presence of the form factor, yields a relation
between the search reaches with and without a running gravitational coupling.
Denoting $M^{FF}$ as the search reach with the form factor, we have
\begin{equation}
M^{\delta+2} =(M^{FF})^{\delta+2}-\Big({{\sqrt s}
\over {t}}\Big)^{\delta+2}\,,
\end{equation}
where here we have chosen $\mu^2=s$ to maximize the effect 
of the form factor. Since values of $M$ are in the 
multi-TeV range, we find that $M^{FF}=M$ to the high accuracy of 
$\sim 0.1\%$ or better for $t \geq 0.5$. 
For example, with $\delta=2~(6)$ the traditional reach is given by 
$M=8.3~(2.9)$ TeV \cite{rev,grw,mp}. 
Taking $t=0.5$ for these cases we obtain $M^{FF}=8.298~(2.898)$ TeV 
when the form factor is present. Choosing a smaller and perhaps more realistic 
value of $\mu$, such as the photon energy, we see that the 
corresponding change in the ILC discovery reach is even further reduced.
Thus to a very good approximation, a running gravitational coupling is seen 
to have very little effect on ILC search reaches for ADD model signatures. 

\section{Warped Extra Dimensions}

In the RS model {\cite {RS}}, the $S^1/Z_2$ orbifolded, slice of 
5-dimensional space bounded by
two branes is described by a non-factorizable metric 
\begin{equation}
ds^2=e^{-2k|y|}\eta_{\mu\nu}dx^\mu dx^\nu -dy^2\,,
\end{equation}
with $\eta_{\mu\nu}$ being the flat Minkowski metric. The two 
branes are separated by a distance $\pi r_c$, with one being located at $y=0$ 
(known as the UV brane), and the other at $y=\pi r_c$ (the IR brane). The warp 
factor, $\epsilon=e^{-\pi kr_c}$, generates the hierarchy between the 
Planck and electroweak scales 
when $kr_c \sim 11$.  The curvature parameter $k$ 
satisfies $k \sim M\sim \overline M_{Pl}$ with $0.01 \lsim 
c=k/\overline M_{Pl} 
\lsim 0.1$ {\cite {us}}.   

In order to keep things simple, we limit ourselves to 
the case where the SM fields are localized to the IR brane;
we can then concentrate on the 
gravitational sector of the theory and readily compare results with the 
classic RS model. In this case, the principle collider signal for the 
RS model is the resonant production of spin-2 KK gravitons \cite{us}.
The KK masses are given by $m_n=x_nk\epsilon$ where $x_n$ are the roots 
of the $J_1$ Bessel function, and they couple to the Standard Model fields 
with electroweak strength.  Since the KK gravitons are directly produced
in the $s$-channel in this scenario, 
the form factor describing the running gravitational coupling
can be written as
\begin{equation}
F^{-1}=1+\Big({\sqrt s\over {tM\epsilon}}\Big)^3\,,
\end{equation}
where we have set $\mu=\sqrt s$.

It is interesting to first consider 
the ratio of width to mass for the graviton KK states as $n$ increases. In 
the standard RS picture,
\begin{equation}
{{\Gamma_n}\over {m_n}}= Nc^2\Big({{m_n}\over {k\epsilon}}\Big)^2=Nc^2x_n^2\,,
\end{equation}
where $N$ is a fixed numerical factor $\simeq 5/16\pi$. 
For a fixed value of $c$ we observe 
that this ratio grows significantly as $n$ increases, and at some point 
the very idea of a graviton resonance is lost. On the resonance peak
for a KK graviton, the form factor given above 
can be written as
\begin{equation}
F^{-1}=1+\Big({{m_n}\over {tM\epsilon}}\Big)^3=1+[x_nc^{2/3}/t]^3\,,
\end{equation}
where the last equality follows from the relation $M^3=k\overline M_{Pl}^2$.
We now see that width scales as 
\begin{equation}
\Gamma_n=Nk\epsilon c^2 x_n^3\Big[1+c^2x_n^3/t^3\Big]^{-1}\,.
\end{equation}
As $x_n$ gets large we now find 
\begin{equation}
\Gamma_n \simeq Nk\epsilon t^3\,,
\end{equation}
which is {\it independent} of both $c$ and $n$. Thus 
the form factor prevents the widths of the KK graviton states 
from growing too large and 
a well-defined resonance structure is maintained for every level
in the KK tower. We also see 
that the parameter $t$ plays an important role in determining 
the graviton width; however, a problem may still arise if the value of 
$t$ is too large. 

\begin{figure}[htbp]
\centerline{
\includegraphics[width=7.0cm,angle=90]{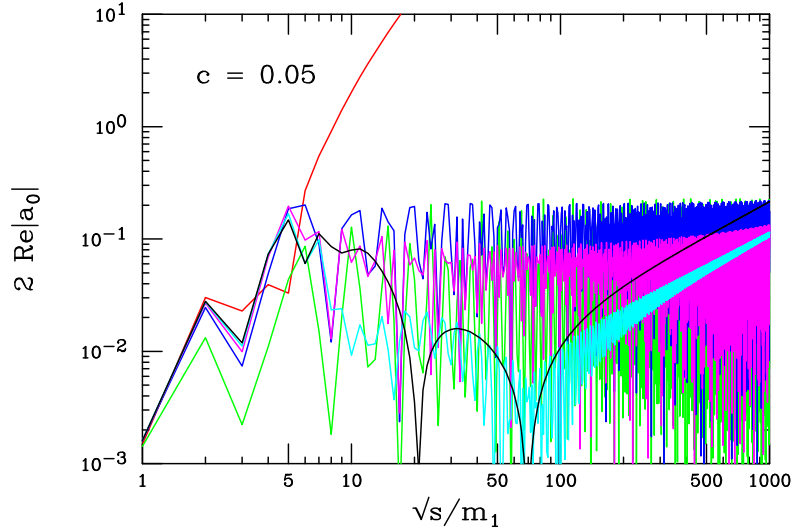}}
\vspace*{0.1cm}
\caption{The $J=0$ partial wave amplitude for
$hh\to hh$ as a function of the ratio $\sqrt s/m_1$, summing over the
first 10,000 KK graviton states in the RS model in the presence of
the form factor.  We set $c=0.05$.  The green, dark blue, magenta, cyan,
and black curves correspond to the values $t=1,~2,~3,~4$ and 5, 
respectively.  The red curve represents the conventional RS rsult.}
\label{rsunit}
\end{figure}

To determine the range of the parameter $t$ that is allowed by
perturbative unitarity in the RS model, we again study the $2\to 2$
scattering process $hh\to hh$ at high energies.  The KK tower of
gravitons contribute to this process via $s$-, $t$- and $u$-channel exchanges.
Including the form factor and summing over the first 10,000 states in
the graviton KK tower, we obtain the results displayed in Fig.~\ref{rsunit}
for the $J=0$ partial wave
amplitude for this channel.  In this figure, we show the
value of $2Re|a_0|$ as a function of the ratio $\sqrt s/m_1$
(where $m_1$ is the mass of the first graviton KK state) for several
values of $t$ assuming $c=0.05$.  It is clear that this amplitude is
well-behaved and $2Re|a_0|$ is always less than unity when $t\leq 2$.
However, at very high energies, it appears that the amplitude starts
to grow for $t\geq 3$ and will eventually violate perturbative
unitarity.  We find that these results do not change appreciably as
$c$ is varied.
Note that the presence of the form factor greatly dampens
this amplitude compared to the standard RS result.
  
\begin{figure}[htbp]
\centerline{
\includegraphics[width=7.0cm,angle=90]{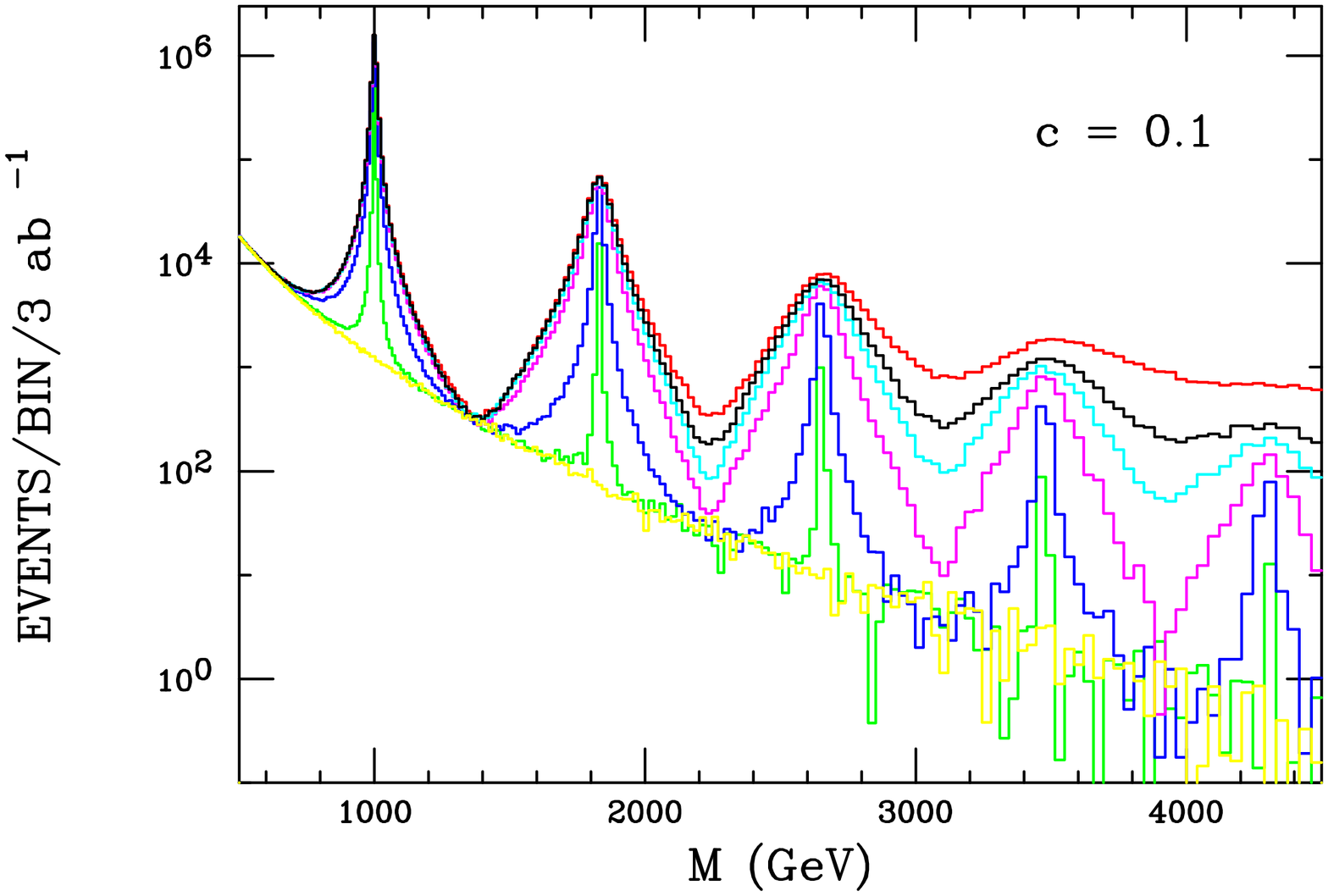}}
\vspace*{0.15cm}
\centerline{
\includegraphics[width=7.0cm,angle=90]{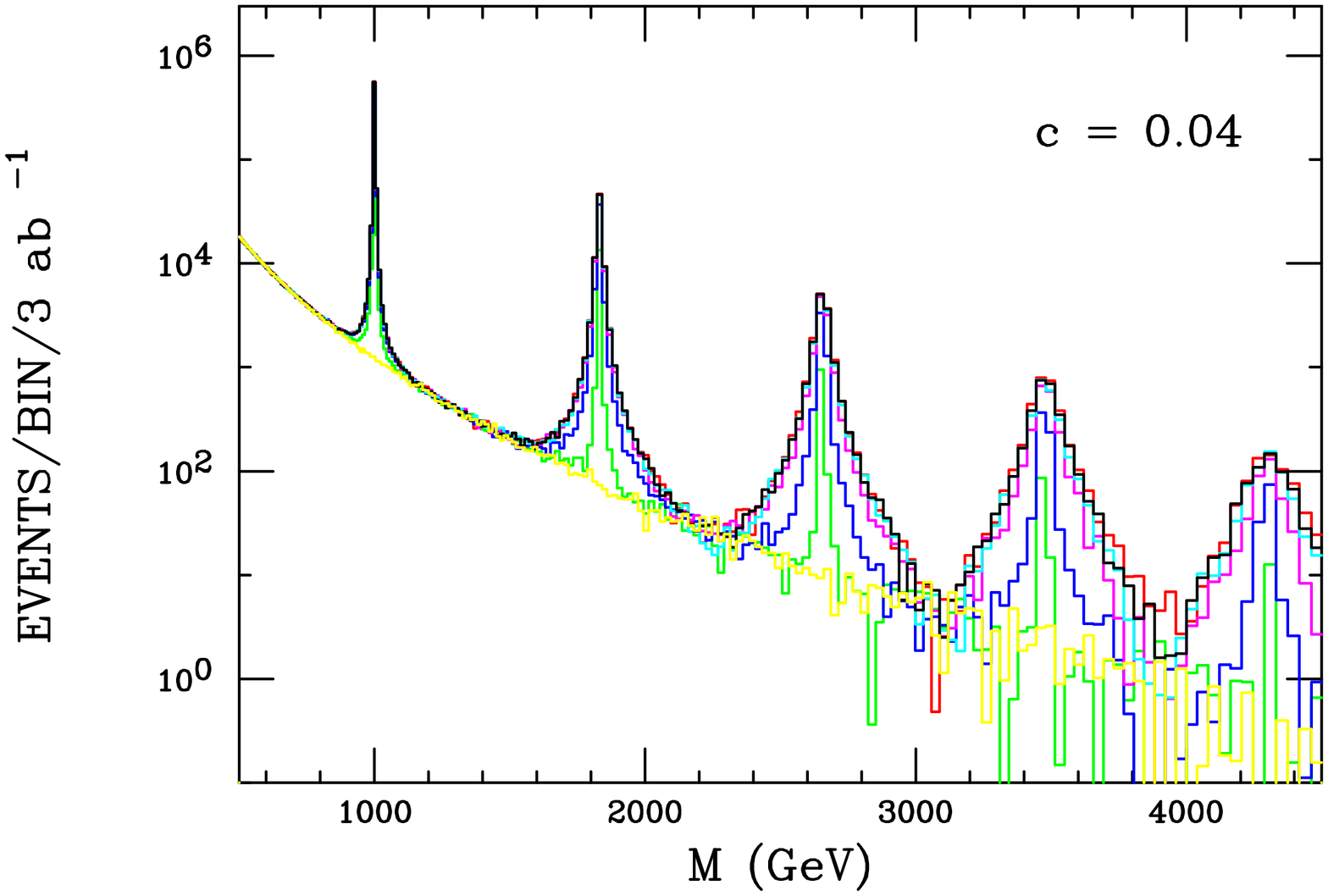}}
\vspace*{0.1cm}
\caption{RS graviton resonance production in the Drell-Yan channel 
as a function 
of the dilepton pair invariant mass at a high luminosity 
LHC assuming $m_1=1$ TeV and $c=0.1(0.04)$ in the upper(lower) panel. 
The lowest(yellow) histogram in the SM background while the 
outermost(red) histogram is the usual RS prediction. From outside 
going inwards the next five histograms correspond to $t=4(3,2,1,0.5)$, 
respectively. In all the form factor cases the scale is assumed to 
be the dilepton pair mass $M$.}
\label{figA}
\end{figure}

The effects of the form factor in the 
production of graviton KK resonances in the Drell-Yan channel at the LHC
are shown in Fig.~\ref{figA}. 
Here we see the familiar pattern that, in the standard RS picture, 
the resonances get wider and wider as the level increases in the KK tower
and the 
resonance structure is completely lost above $n=3-4$ depending 
upon the value of $c$. Turning on the form factor and taking smaller 
values of $t$, we do not lose too much of the apparent signal peak,
but the towers separate and become more narrow for 
large $n$. Certainly, for $t\lsim 2$ the resonance structure is always 
quite clean for the range of KK tower masses shown in the Figure. 
Note that if RS graviton KK resonances are observed and such form factors 
are present, the value of $t$ will be relatively easy to 
extract from the cross section data.

\begin{figure}[htbp]
\centerline{
\includegraphics[width=7.5cm,angle=90]{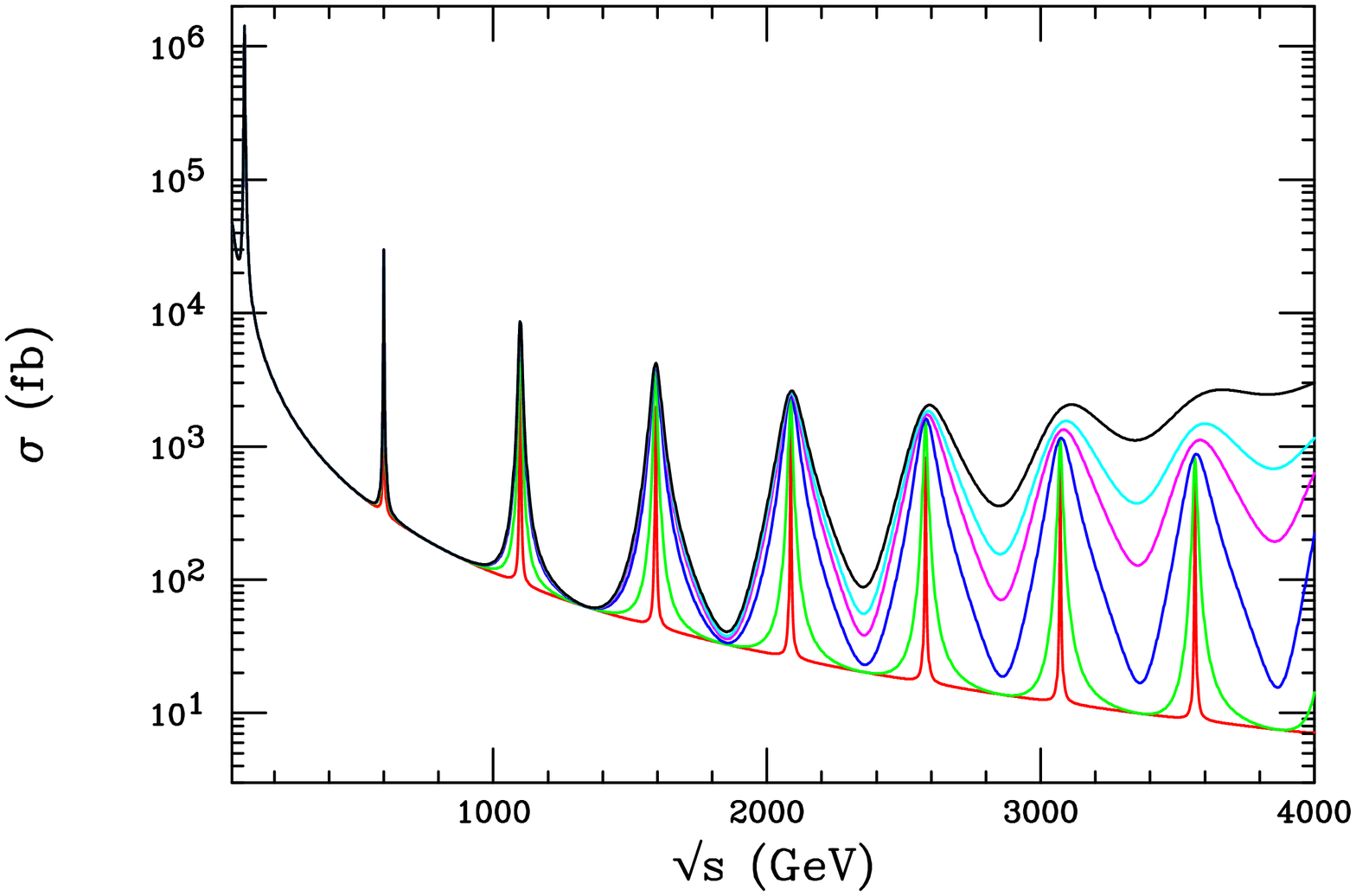}}
\vspace*{0.15cm}
\centerline{
\includegraphics[width=7.5cm,angle=90]{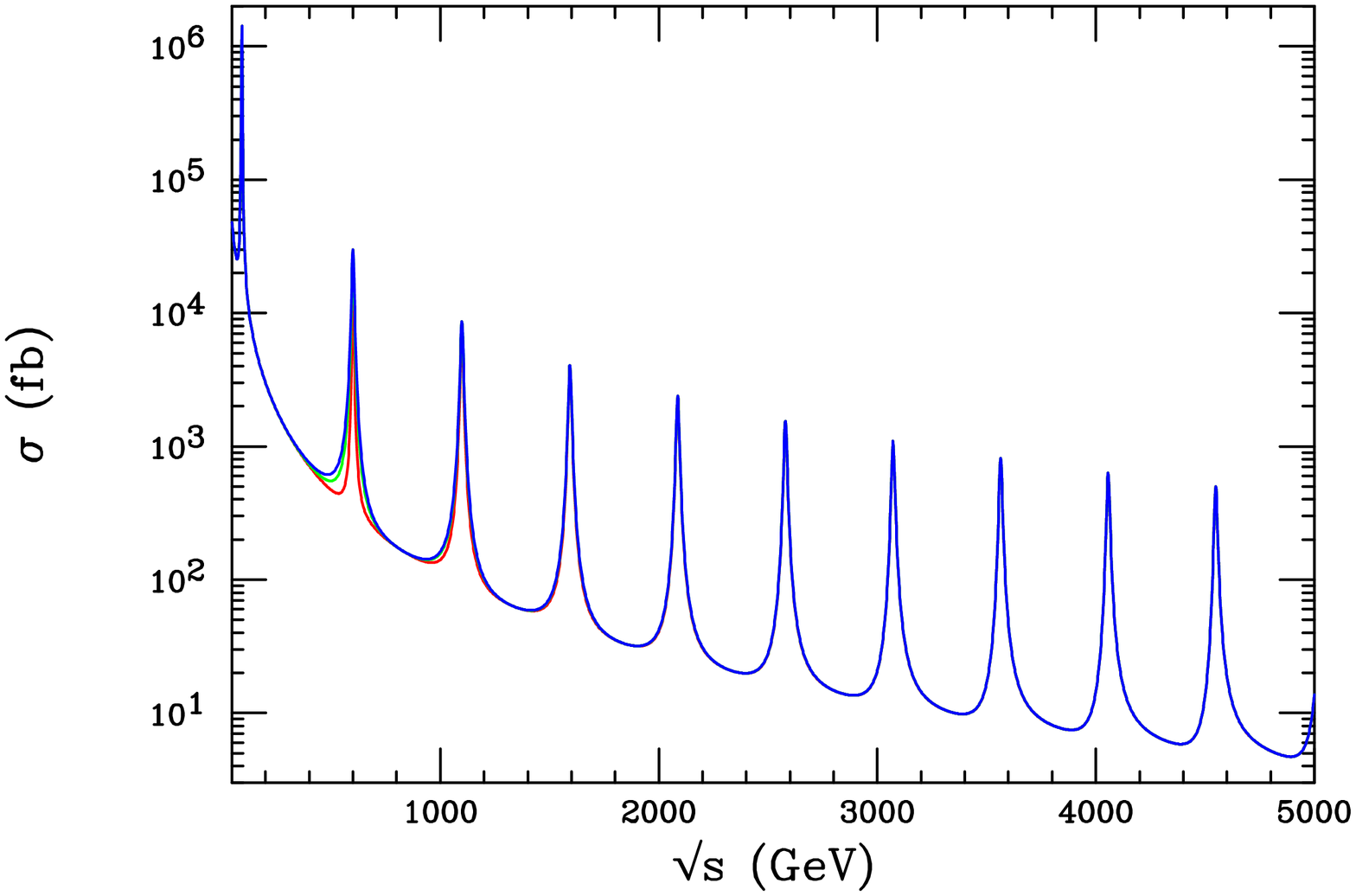}}
\vspace*{0.1cm}
\caption{Cross section for $e^+e^- \to \mu^+\mu^-$ as a function 
of $\sqrt s$ in the RS model with $c=0.05$ and $m_1=$600 GeV showing 
the first 7(9) graviton KK excitations in the top(bottom) panel. 
In the top panel, 
the uppermost curve corresponds to the standard result. 
Moving inward the curves correspond to the case where a form factor 
is present assuming $t=4(3,2,1,0.5)$, respectively. In the lower 
panel, $t=1$ has been assumed for very large values of $c=0.1,0.2$ 
and 0.3, with the narrowest resonance curve corresponding to the 
smallest value of $c$.}
\label{figB}
\end{figure}

The effect of the form factor on the widths of the KK graviton 
resonances become even more obvious at multi-TeV scale 
$e^+e^-$ colliders such as CLIC \cite{Accomando:2004sz}. 
Figure~\ref{figB} shows the 
cross section for the process  $e^+e^- \to \mu^+\mu^-$ as a function of 
$\sqrt s$ with $m_1=600$ GeV. In the upper panel we assume $c=0.05$ 
and see the loss of resonance structure and the potential 
unitarity violation in the conventional RS model and how this 
situation is tamed by the presence of a form factor. Certainly for 
$t \lsim 2$ we see that the narrow resonance structure is maintained for
all levels of the KK tower.  In the 
lower panel, for fixed $t=1$, it is clear that the graviton KK 
resonances are all quite narrow and are essentially $c$ and 
$n$ independent, even for high levels in the KK tower.
Only for the lightest KK 
state do we see dependence on $c$ and even in this case it is 
rather weak. From these figures we can see that the form factor 
modifies the cross section as advertised.

\section{Summary and Discussion}

The poor high energy behavior of General Relativity observed in perturbation
theory may be cured if there exists a non-Gaussian fixed point rendering the
theory asymptotically safe and potentially non-perturbatively renormalizable.
If such a possibility is realized, the effective gravitational coupling at
high energies becomes weaker and this running can be parameterized through
the introduction of a form factor when calculating the interactions of
gravitons with matter or each other. These form factors can also modify
graviton exchange amplitudes rendering them unitary at tree-level. The
evidence that such a situation may be realized in nature is reasonably
strong and has improved theoretically in recent years. However, since the
effect of the form factor is only significant once the relevant energies
approach the fundamental scale of gravity, 
in 4-dimensions it will be difficult to test this
possibility directly anytime in the near future.
               
Extra-dimensional scenarios of the ADD or RS type allow for the fundamental
scale to be not far above $\sim 1$ TeV. If such scenarios are realized, the
existence of gravitational form factors can then be probed at future colliders
such as the LHC and/or the ILC. In this paper we have shown that this is
indeed the case for both these scenarios.
                                        
The analysis of Reuter \etal\ and of Litim \cite{bigref} 
suggests a very specific
structure for this form factor which is totally determined in D-dimensions
up to an order one
coefficient, $t$. We find that imposing tree-level unitarity requirements
on graviton exchange amplitudes in either the ADD or RS models implies that
the range of $t$ is restricted: $t \lsim 2$. In both of these models,
graviton exchange processes were shown to be particularly sensitive to the
presence of these form factors. In particular, we demonstrated that
measurements at both the LHC and ILC can be used to extract the value of $t$.
In the RS model, the
width of graviton resonances, which ordinarily increases at higher levels of
the KK tower, was shown to asymptote to a constant value when form factors are
employed. On the otherhand, the process of graviton emission
which occurs in the ADD
scenario, was shown to be rather insensitive to the presence of form factors.
Interestingly, the collider search reaches for extra dimension was also shown
to not be overly sensitive to form factor contributions in both the ADD and RS
cases.

If these extra dimensional scenarios are realized in Nature, then    
hopefully the observation of form factor effects 
will be observed, thus providing us with an important handle on the underlying
theory of quantum gravity.

%
\def\MPL #1 #2 #3 {Mod. Phys. Lett. {\bf#1},\ #2 (#3)}
\def\NPB #1 #2 #3 {Nucl. Phys. {\bf#1},\ #2 (#3)}
\def\PLB #1 #2 #3 {Phys. Lett. {\bf#1},\ #2 (#3)}
\def\PR #1 #2 #3 {Phys. Rep. {\bf#1},\ #2 (#3)}
\def\PRD #1 #2 #3 {Phys. Rev. {\bf#1},\ #2 (#3)}
\def\PRL #1 #2 #3 {Phys. Rev. Lett. {\bf#1},\ #2 (#3)}
\def\RMP #1 #2 #3 {Rev. Mod. Phys. {\bf#1},\ #2 (#3)}
\def\NIM #1 #2 #3 {Nuc. Inst. Meth. {\bf#1},\ #2 (#3)}
\def\ZPC #1 #2 #3 {Z. Phys. {\bf#1},\ #2 (#3)}
\def\EJPC #1 #2 #3 {E. Phys. J. {\bf#1},\ #2 (#3)}
\def\IJMP #1 #2 #3 {Int. J. Mod. Phys. {\bf#1},\ #2 (#3)}
\def\JHEP #1 #2 #3 {J. High En. Phys. {\bf#1},\ #2 (#3)}


\begin{thebibliography}{99}

\bibitem{sm}
J.~Alcaraz {\it et al.}  [ALEPH Collaboration],
  arXiv:hep-ex/0612034;
S.~Bethke,
  Prog.\ Part.\ Nucl.\ Phys.\  {\bf 58}, 351 (2007)
  [arXiv:hep-ex/0606035];
 W.~M.~Yao {\it et al.}  [Particle Data Group],
  J.\ Phys.\ G {\bf 33} (2006) 1;
J.~L.~Hewett,
  ``The standard model and why we believe it,'' Presented at {\it TASI97:
Supersymmetry, Supergravity, and Supercolliders}, Boulder, CO, June 1997,
  arXiv:hep-ph/9810316.

\bibitem{adel}
D.~J.~Kapner, T.~S.~Cook, E.~G.~Adelberger, J.~H.~Gundlach,
 B.~R.~Heckel, C.~D.~Hoyle and H.~E.~Swanson,
  Phys.\ Rev.\ Lett.\  {\bf 98}, 021101 (2007)
  [arXiv:hep-ph/0611184];
E.~G.~Adelberger, B.~R.~Heckel, S.~Hoedl, C.~D.~Hoyle, 
D.~J.~Kapner and A.~Upadhye,
  Phys.\ Rev.\ Lett.\  {\bf 98}, 131104 (2007)
  [arXiv:hep-ph/0611223].

\bibitem{solar}
See, for example,
  I.~Navarro and K.~Van Acoleyen,
  arXiv:gr-qc/0512109 and
  arXiv:gr-qc/0511045,
  Phys.\ Lett.\ B {\bf 622}, 1 (2005)
  [arXiv:gr-qc/0506096];
  G.~J.~Olmo,
  Phys.\ Rev.\ Lett.\  {\bf 95}, 261102 (2005)
  [arXiv:gr-qc/0505101] and
  Phys.\ Rev.\ D {\bf 72}, 083505 (2005);
  M.~T.~Jaekel and S.~Reynaud,
  Mod.\ Phys.\ Lett.\ A {\bf 20}, 1047 (2005)
  [arXiv:gr-qc/0410148];
  A.~D.~Dolgov and M.~Kawasaki,
  Phys.\ Lett.\ B {\bf 573}, 1 (2003)
  [arXiv:astro-ph/0307285];
  T.~Chiba,
  Phys.\ Lett.\ B {\bf 575}, 1 (2003)
  [arXiv:astro-ph/0307338];
  M.~E.~Soussa and R.~P.~Woodard,
  Gen.\ Rel.\ Grav.\  {\bf 36}, 855 (2004)
  [arXiv:astro-ph/0308114];
  S.~Capozziello and A.~Troisi,
  Phys.\ Rev.\ D {\bf 72}, 044022 (2005)
  [arXiv:astro-ph/0507545].;
  S.~Capozziello,
  arXiv:gr-qc/0412088;
  I.~Quandt and H.~J.~Schmidt,
  Astron.\ Nachr.\  {\bf 312}, 97 (1991)
  [arXiv:gr-qc/0109005];
  S.~Nojiri and S.~D.~Odintsov,
  Phys.\ Rev.\ D {\bf 68}, 123512 (2003)
  [arXiv:hep-th/0307288];
  M.~C.~B.~Abdalla, S.~Nojiri and S.~D.~Odintsov,
  Class.\ Quant.\ Grav.\  {\bf 22}, L35 (2005)
  [arXiv:hep-th/0409177].
See, however,
  K.~Tangen,
  arXiv:gr-qc/0602089.

\bibitem{cosmo}
See, for example,
  S.~M.~Carroll, A.~De Felice, V.~Duvvuri, D.~A.~Easson, M.~Trodden 
and M.~S.~Turner,
  Phys.\ Rev.\ D {\bf 71}, 063513 (2005)
  [arXiv:astro-ph/0410031];
  S.~M.~Carroll, V.~Duvvuri, M.~Trodden and M.~S.~Turner,
  Phys.\ Rev.\ D {\bf 70}, 043528 (2004)
  [arXiv:astro-ph/0306438];
  S.~Nojiri and S.~D.~Odintsov,
  arXiv:hep-th/0601213;
  T.~Clifton and J.~D.~Barrow,
  arXiv:gr-qc/0601118;
  N.~Arkani-Hamed, S.~Dimopoulos, G.~Dvali and G.~Gabadadze,
  arXiv:hep-th/0209227;
  T.~Biswas, A.~Mazumdar and W.~Siegel,
  arXiv:hep-th/0508194;
  G.~R.~Dvali, G.~Gabadadze and M.~Porrati,
  Phys.\ Lett.\ B {\bf 485}, 208 (2000)
  [arXiv:hep-th/0005016];
  G.~Kofinas, R.~Maartens and E.~Papantonopoulos,
  JHEP {\bf 0310}, 066 (2003)
  [arXiv:hep-th/0307138];
  R.~A.~Brown, R.~Maartens, E.~Papantonopoulos and V.~Zamarias,
  JCAP {\bf 0511}, 008 (2005)
  [arXiv:gr-qc/0508116];
  P.~D.~Mannheim and D.~Kazanas,
  Astrophys.\ J.\  {\bf 342}, 635 (1989);
  H.~J.~Schmidt,
  Int.\ J.\ Mod.\ Phys.\ A {\bf 5}, 4661 (1990);
  T.~Biswas, A.~Mazumdar and W.~Siegel,
  arXiv:hep-th/0508194;
  S.~Nojiri and S.~D.~Odintsov,
  arXiv:hep-th/0601213 and
  S.~Nojiri and S.~D.~Odintsov,
  Phys.\ Lett.\ B {\bf 631}, 1 (2005)
  [arXiv:hep-th/0508049];
  G.~Cognola, E.~Elizalde, S.~Nojiri, S.~D.~Odintsov and S.~Zerbini,
  arXiv:hep-th/0601008;
  M.~Amarzguioui, O.~Elgaroy, D.~F.~Mota and T.~Multamaki,
  arXiv:astro-ph/0510519.
  


\bibitem{Feynman:1963ax}
  R.~P.~Feynman,
  Acta Phys.\ Polon.\  {\bf 24}, 697 (1963);
  R.~P.~Feynman, F.~B.~Morinigo, W.~G.~Wagner and B.~Hatfield,
{SPIRES entry}
{\it  Reading, USA: Addison-Wesley (1995) 232 p. (The advanced book program).}

\bibitem{Ward:2006bc}
  B.~F.~L.~Ward,
  arXiv:hep-ph/0607198;
  B.~F.~L.~Ward,
  Acta Phys.\ Polon.\  B {\bf 37}, 1967 (2006)
  [arXiv:hep-ph/0605054].

\bibitem{Weinberg}
S. Weinberg in {\it General Relativity}, eds. S.W. Hawking and W. Israel, 
(Cambridge 
Univ. Press, Cambridge, 1979) p.790.

\bibitem{strings}
See, for example, M. Green, J. Schwarz and E. Witten, {\it Superstring 
Theory, v.1 and v.2}, (Cambbridge Univ. Press, Cambridge, 1987); and J.
Polchinski, {\it String Theory, v.1 and 
v.2}, (Cambridge Univ. Press, Cambridge, 1998). 

\bibitem{Smolin:2004sx}
  L.~Smolin,
  arXiv:hep-th/0408048.

\bibitem{compgrav}
See, for example,   T.~Okui,
  Phys.\ Rev.\  D {\bf 73}, 075012 (2006)
  [arXiv:hep-ph/0511082];
  D.~L.~Henty,
  arXiv:hep-th/0312255;
  G.~Dvali,
  New J.\ Phys.\  {\bf 8}, 326 (2006)
  [arXiv:hep-th/0610013].


\bibitem{Yennie:1961ad}
  D.~R.~Yennie, S.~C.~Frautschi and H.~Suura,
  Annals Phys.\  {\bf 13}, 379 (1961).

\bibitem{bigref}
  O.~Lauscher and M.~Reuter,
  arXiv:hep-th/0511260;
  O.~Lauscher and M.~Reuter,
  Phys.\ Rev.\  D {\bf 66}, 025026 (2002);
  M.~Reuter and F.~Saueressig,
  Phys.\ Rev.\  D {\bf 65}, 065016 (2002)
  [arXiv:hep-th/0110054];
  O.~Lauscher and M.~Reuter,
  Class.\ Quant.\ Grav.\  {\bf 19}, 483 (2002)
  [arXiv:hep-th/0110021];
  O.~Lauscher and M.~Reuter,
  Phys.\ Rev.\  D {\bf 65}, 025013 (2002)
  [arXiv:hep-th/0108040];
 A.~Codello and R.~Percacci,
 Phys.\ Rev.\ Lett.\  {\bf 97}, 221301 (2006)
 [arXiv:hep-th/0607128];
  D.~F.~Litim,
  Phys.\ Rev.\ Lett.\  {\bf 92}, 201301 (2004)
  [arXiv:hep-th/0312114];
  H.~Emoto,
  arXiv:gr-qc/0612127.

\bibitem{Hamber:2006sv}
  H.~W.~Hamber and R.~M.~Williams,
  Phys.\ Rev.\  D {\bf 75}, 084014 (2007)
  [arXiv:hep-th/0607228].

\bibitem{Fischer:2006at}
  P.~Fischer and D.~F.~Litim,
  AIP Conf.\ Proc.\  {\bf 861}, 336 (2006)
  [arXiv:hep-th/0606135].

\bibitem{ADD}
N.~Arkani-Hamed, S.~Dimopoulos and G.~R.~Dvali,
Phys.\ Rev.\ D {\bf 59}, 086004 (1999)
[arXiv:hep-ph/9807344] and
Phys.\ Lett.\ B {\bf 429}, 263 (1998)
[arXiv:hep-ph/9803315];
I.~Antoniadis, N.~Arkani-Hamed, S.~Dimopoulos and G.~R.~Dvali,
Phys.\ Lett.\ B {\bf 436}, 257 (1998)
[arXiv:hep-ph/9804398].

\bibitem{RS}
L.~Randall and R.~Sundrum,
Phys.\ Rev.\ Lett.\  {\bf 83}, 3370 (1999)
[arXiv:hep-ph/9905221].

\bibitem{Niedermaier:2006ns}
  M.~Niedermaier,
  arXiv:gr-qc/0610018.

\bibitem{Bonanno:2006eu}
  A.~Bonanno and M.~Reuter,
  Phys.\ Rev.\  D {\bf 73}, 083005 (2006)
  [arXiv:hep-th/0602159].

\bibitem{rev}
For a review, see 
J.~Hewett and M.~Spiropulu,
Ann.\ Rev.\ Nucl.\ Part.\ Sci.\  {\bf 52}, 397 (2002)
[arXiv:hep-ph/0205106].

\bibitem{me}
 J.~L.~Hewett,
  Phys.\ Rev.\ Lett.\  {\bf 82}, 4765 (1999)
  [arXiv:hep-ph/9811356];
  T.~G.~Rizzo,
  Phys.\ Rev.\  D {\bf 59}, 115010 (1999)
  [arXiv:hep-ph/9901209].

\bibitem{grw}
  G.~F.~Giudice, R.~Rattazzi and J.~D.~Wells,
  Nucl.\ Phys.\  B {\bf 544}, 3 (1999)
  [arXiv:hep-ph/9811291].

\bibitem{He:1999qv}
  X.~G.~He,
  Phys.\ Rev.\  D {\bf 61}, 036007 (2000)
  [arXiv:hep-ph/9905500].

\bibitem{mp}
  E.~A.~Mirabelli, M.~Perelstein and M.~E.~Peskin,
  Phys.\ Rev.\ Lett.\  {\bf 82}, 2236 (1999)
  [arXiv:hep-ph/9811337].

\bibitem{Vacavant:2001sd}
  L.~Vacavant and I.~Hinchliffe,
  J.\ Phys.\ G {\bf 27}, 1839 (2001).

\bibitem{teslatdr}
  J.~A.~Aguilar-Saavedra {\it et al.}  [ECFA/DESY LC Physics Working Group],
  arXiv:hep-ph/0106315.

\bibitem{us}
  H.~Davoudiasl, J.~L.~Hewett and T.~G.~Rizzo,
  Phys.\ Rev.\ Lett.\  {\bf 84}, 2080 (2000)
  [arXiv:hep-ph/9909255].

\bibitem{Accomando:2004sz}
  E.~Accomando {\it et al.}  [CLIC Physics Working Group],
  arXiv:hep-ph/0412251.



\end{thebibliography}
\end{document}